\documentclass[11pt]{article}
\usepackage{epsfig}

\newcommand{\sect}[1]{ \section{#1} \setcounter{equation}{0} }

\newcommand{\MSbar}{\overline{\mbox{MS}}}

\begin{document}
\date{}
\title{\textbf{Remarks on a class of renormalizable interpolating gauges }}
\author{\textbf{D. Dudal}$^{a}$\thanks{%
Research Assistant of The Fund For Scientific Research-Flanders,
Belgium.} \
, \textbf{J.A. Gracey}$^{b}$\thanks{%
jag@amtp.liv.ac.uk } \ , \textbf{V.E.R. Lemes}$^{c}$\thanks{%
vitor@dft.if.uerj.br} \ , \textbf{R.F. Sobreiro}$^{c}$\thanks{%
sobreiro@uerj.br} \and \textbf{S.P. Sorella}$\thanks{%
sorella@uerj.br}{\ }\footnote{Work supported by FAPERJ, Funda{\c
c}{\~a}o de Amparo {\`a} Pesquisa do Estado do Rio de Janeiro,
under the program {\it Cientista do Nosso Estado}, E-26/151.947/2004.}^{c}$\thanks{%
sorella@uerj.br} \ , \textbf{R. Thibes}$^{c}$\thanks{%
thibes@dft.if.uerj.br} \ , \textbf{H. Verschelde}$^{a}$\thanks{%
david.dudal@ugent.be, henri.verschelde@ugent.be} \\
\\
\textit{$^{a}$ Ghent University} \\
\textit{Department of Mathematical Physics and Astronomy} \\
\textit{Krijgslaan 281-S9, B-9000 Gent, Belgium}\\
[3mm] \textit{$^{b}$ Theoretical Physics Division} \\
\textit{Department of Mathematical Sciences} \\
\textit{University of Liverpool} \\
\textit{P.O. Box 147, Liverpool, L69 3BX, United Kingdom} \\
[3mm] \textit{$^{c}$ UERJ, Universidade do Estado do Rio de Janeiro} \\
\textit{Rua S{\~a}o Francisco Xavier 524, 20550-013 Maracan{\~a}} \\
\textit{Rio de Janeiro, Brasil}} \maketitle

\begin{abstract}
\noindent A class of covariant gauges allowing one to interpolate
between the Landau, the maximal Abelian, the linear covariant and
the Curci-Ferrari gauges is discussed. Multiplicative
renormalizability is proven to all orders by means of algebraic
renormalization. All one-loop anomalous dimensions of the fields and
gauge parameters are explicitly evaluated in the $\MSbar$ scheme.
\end{abstract}
\vspace{-16cm} \hfill LTH--652 \vspace{16cm}
\newpage

\sect{Introduction.} Interpolating gauges have been proven to be a
powerful tool to achieve a better understanding of the behavior of
the Green's functions in different gauges. Let us quote, for
instance, the interpolating gauge introduced in
\cite{Piguet:1989va}, allowing one to connect the Landau and the
light cone gauges, thereby providing a useful framework to
investigate Yang-Mills theories in noncovariant gauges. Another
example is provided by the interpolating gauge proposed in
\cite{Baulieu:1998kx,Fischer:2005qe}, which connects the Landau and
the Coulomb gauges. More recently, a class of covariant gauges
interpolating between the Landau and the maximal Abelian gauge has
been introduced in \cite{A3} in order to investigate the properties
of the vacuum energy of Yang-Mills theories due to dimension two
gauge condensates. More precisely, it has been possible to show that
the vacuum energy obtained in the maximal Abelian gauge, due to
the condensate $\langle \frac{A_{\mu }^{a}A_{\mu }^{a}}{2}+\alpha \overline{c%
}^{a}c^{a}\rangle $, is the same as the vacuum energy obtained in
the Landau
gauge, due to the condensate\footnote{In the case of the maximal Abelian gauge, the index $a$ refers only to the $%
N(N-1)$ off-diagonal generators of $SU(N)$, while the index $A$ runs
from 1 to $N^2-1$.} $\langle \frac{A_{\mu }^{A}A_{\mu }^{A}}{2}%
\rangle$. We recall that a nonvanishing condensate $\langle
\frac{A_{\mu }^{a}A_{\mu }^{a}}{2}+\alpha
\overline{c}^{a}c^{a}\rangle $ in the maximal Abelian gauge results
in a dynamical mass generation for the off-diagonal gluons. This can
be seen as evidence for the Abelian dominance hypothesis, implying
that Yang-Mills theories in the low energy region should be
described by an effective abelian theory, according to the dual
superconductivity mechanism for color confinement.

In this work we introduce a class of covariant gauges allowing one
to interpolate between the Landau gauge, the maximal Abelian gauge,
the linear covariant gauges and the Curci-Ferrari gauge. It is worth
remarking that in all these gauges a nonvanishing dimension two
condensate has emerged (see \cite{A3,Dudal:2003by} and references
therein), so that this generalized interpolating gauge might be
useful for further investigation of these condensates.

The work is organized as follows. In Sect. 2 we introduce the
interpolating
gauge fixing. Since the maximal Abelian gauge requires a splitting of the $%
SU(N)$ color index $A=(i,a)$, corresponding to the $N-1$ diagonal
generators
of the Cartan subalgebra of $SU(N)$, labeled by the index $i$, and to the $%
N(N-1)$ off-diagonal generators, labeled by $a$, it turns out that
the gauge-fixing term contains six independent gauge parameters. In
Sect. 3 we discuss the various limits for those gauge parameters,
allowing one to recover the linear covariant gauges, the Landau
gauge, the maximal Abelian gauge and the Curci-Ferrari gauge. In
Sect. 4 we prove the all orders multiplicative renormalizability by
means of the algebraic renormalization procedure. In Sect. 5 we
present the explicit one-loop calculation of the anomalous
dimensions of all fields and parameters. These reduce to the already
known values in the various limits for the interpolating gauge
parameters. In particular, as a useful check, it is verified that
the beta function of the gauge coupling is independent of all the
gauge parameters. In the concluding section we underline a potential
application of the interpolating gauge for lattice simulations. Some
technical details are collected in Appendix A.

\sect{The interpolating gauge fixing term.}
Let $A_{\mu }$ be the Lie algebra valued connection for the gauge group $%
SU(N)$, whose generators $T^{A}$, satisfying $\left[ T^{A},T^{B}\right] $~$=$%
~$if^{ABC}T^{C}$, are chosen to be hermitean and to obey the
orthonormality
condition $\mathrm{Tr}\left( T^{A}T^{B}\right) =T_{F}\delta ^{AB}$, with $%
A,B,C=1,\ldots ,\left( N^{2}-1\right) $. As needed in order to
discuss the maximal Abelian gauge, we decompose the gauge field into
off-diagonal and diagonal components, namely
\begin{equation}
A_{\mu }=A_{\mu }^{A}T^{A}=A_{\mu }^{a}T^{a}+A_{\mu }^{i}T^{\,i},
\label{conn}
\end{equation}
where the indices $i$, $j$, $k$ label the $N-1$ generators of the
Cartan subalgebra. The remaining $N(N-1)$ off-diagonal generators
will be labelled by the rest of the small Latin indices. For later
use, we recall the Jacobi identity
\begin{equation}
f^{ABC}f^{CDE}+f^{ADC}f^{CEB}+f^{AEC}f^{CBD}=0\;,  \label{jacobi}
\end{equation}
from which it can be deduced that
\begin{eqnarray}
f^{abi}f^{bjc}+f^{abj}f^{bci} &=&0\;,  \label{jacobi2} \\
f^{abc}f^{bdi}+f^{abd}f^{bic}+f^{abi}f^{bcd} &=&0\;,  \nonumber \\
f^{abc}f^{cde}+f^{abi}f^{ide}+f^{adc}f^{ceb}+f^{adi}f^{ieb}+f^{aec}f^{cbd}+f^{aei}f^{ibd}
&=&0\;. \nonumber
\end{eqnarray}
The field strength decomposes as
\begin{equation}
F_{\mu \nu }=F_{\mu \nu }^{A}T^{A}=F_{\mu \nu }^{a}T^{a}+F_{\mu \nu
}^{i}T^{\,i}\;,  \label{fs}
\end{equation}
with the off-diagonal and diagonal parts given respectively by
\begin{eqnarray}
F_{\mu \nu }^{a} &=&D_{\mu }^{ab}A_{\nu }^{b}-D_{\nu }^{ab}A_{\mu
}^{b}\;+g\,f^{abc}A_{\mu }^{b}A_{\nu }^{c}\;,  \label{fsc} \\
F_{\mu \nu }^{i} &=&\partial _{\mu }A_{\nu }^{i}-\partial _{\nu
}A_{\mu }^{i}+gf^{abi}A_{\mu }^{a}A_{\nu }^{b}\;,  \nonumber
\end{eqnarray}
where the covariant derivative $D_{\mu }^{ab}$ is defined with
respect to the diagonal components $A_{\mu }^{i}$
\begin{equation}
D_{\mu }^{ab}\equiv \partial _{\mu }\delta ^{ab}-gf^{abi}A_{\mu
}^{i} \;.  \label{cv}
\end{equation}
For the Yang-Mills action one obtains
\begin{equation}
S_{\mathrm{YM}}=-\frac{1}{4}\int d^{4}x\,\left( F_{\mu \nu
}^{a}F^{\mu \nu a}+F_{\mu \nu }^{i}F^{\mu \nu i}\right) \;.
\label{ym}
\end{equation}
\noindent In order to write down the interpolating gauge, we shall
consider the most general gauge fixing term compatible with the
decomposition of the gauge field into diagonal and off-diagonal
components, according to eq.$\left( \ref {conn}\right) $. Let us
first establish the nilpotent BRST transformations of the fields.
From
\begin{eqnarray}
sA_{\mu }^{A} &=&-D_{\mu }^{AB}c^{B}=-\left( \partial _{\mu
}c^{A}+gf^{\,ABC}A_{\mu }^{B}c^{C}\right) \;,  \nonumber \\
sc^{A} &=&\frac{g}{2}f^{ABC}c^{B}c^{C}\;,  \nonumber \\
s\bar{c}^{A} &=&b^{A}\;,  \nonumber \\
sb^{A} &=&0\;,\;  \label{brst0}
\end{eqnarray}
one easily gets
\begin{eqnarray}
sA_{\mu }^{a} &=&-\left( D_{\mu }^{ab}c^{b}+gf^{\,abc}A_{\mu
}^{b}c^{c}+gf^{\,abi}A_{\mu }^{b}c^{i}\right) ,\,\,\,\;sA_{\mu
}^{i}=-\left(
\partial _{\mu }c^{i}+gf\,^{iab}A_{\mu }^{a}c^{b}\right) \;,  \nonumber \\
sc^{a} &=&gf\,^{abi}c^{b}c^{i}+\frac{g}{2}f\,^{abc}c^{b}c^{c},\,\,\,\,\,\,\,%
\,\,\,\,\,\,\,\,\,\,\,\,\,\,\,\,\,\,\,\,\,\,\,\,\,\,\,\,\,\,\,\,\,\,\,sc^{i}=%
\frac{g}{2}\,f\,^{iab}c^{a}c^{b},  \nonumber \\
s\overline{c}^{a}
&=&b^{a}\;,\,\,\,\,\,\,\,\,\,\,\,\,\,\,\,\,\,\,\,\,\,\,\,\,\,\,\,\,\,\,\,\,%
\,\,\,\,\,\,\,\,\,\,\,\,\,\,\,\,\,\,\,\,\,\,\,\,\,\,\,\,\,\,\,\,\,\,\,\,\,\,%
\,\,\,\,\,\,\,\,\,\,\,\,\,\,\,\,\,\,\,\,\,\,s\overline{c}^{i}=b^{i}\;,
\nonumber \\
sb^{a}
&=&0\;,\,\,\,\,\,\,\,\,\,\,\,\,\,\,\,\,\,\,\,\,\,\,\,\,\,\,\,\,\,\,\,\,\,\,%
\,\,\,\,\,\,\,\,\,\,\,\,\,\,\,\,\,\,\,\,\,\,\,\,\,\,\,\,\,\,\,\,\,\,\,\,\,\,%
\,\,\,\,\,\,\,\,\,\,\,\,\,\,\,\,\,\,\,\,\,\,sb^{i}=0\;. \label{brst}
\end{eqnarray}
Here $(c^{a},c^{i})$ are the off-diagonal and the diagonal
components of the
Faddeev-Popov ghost field, while $(\overline{c}^{a},b^{a})$ and $(\overline{c%
}^{i},b^{i})$ are the off-diagonal and diagonal antighosts and
Lagrange
multipliers, respectively. We also observe that the BRST\ transformations $%
\left( \ref{brst}\right) $ have been obtained by their standard form
upon projection on the off-diagonal and diagonal components of the
fields.

Thus, for the most general BRST\ invariant gauge fixing term we
write
\begin{eqnarray}
S_{\mathrm{GF}} &=&s\,\int d^{4}x\left[ \overline{c}^{a}\left( \partial ^{\mu }{A}%
_{\mu }^{a}+\frac{\alpha _{1}}{2}b^{a}+\alpha _{2}gf^{abi}A_{\mu
}^{b}A^{i\mu }\right) +\bar{c}^{i}\left( \partial ^{\mu }{A}_{\mu }^{i}+%
\frac{\alpha _{3}}{2}b^{i}\right) \right.  \nonumber \\
 &+&\left.\frac{\alpha _{4}}{4}gf^{abc}c^{a}\bar{c}^{b}\bar{c}^{c}+\frac{%
\alpha _{5}}{4}gf^{abi}\bar{c}^{a}\bar{c}^{b}c^{i}+\frac{\alpha _{6}}{2}%
gf^{abi}c^{a}\bar{c}^{b}\bar{c}^{i}\right] \;,  \label{gf}
\end{eqnarray}
where $\alpha _{1},\;\alpha _{2},\;\alpha _{3},\;\alpha
_{4},\;\alpha _{5},\;\alpha _{6}$ are six independent parameters. It
is not difficult to see that the gauge fixing (\ref{gf})
interpolates between several gauges, for instance: the linear
covariant gauges, including the Landau gauge as a special case, the
maximal Abelian gauge, the modified maximal Abelian gauge and the
Curci-Ferrari gauge. This will be checked in the next section, where
the various limits for the gauge parameters will be analyzed in detail.%

The expression (\ref{gf}) is easily worked out and yields
\begin{eqnarray}
S_{\mathrm{GF}} &=&\int d^{4}x\left[ b^{a}\left( \partial ^{\mu }{A}_{\mu }^{a}+\frac{%
\alpha _{1}}{2}b^{a}+\alpha _{2}gf^{abi}A_{\mu }^{b}A^{i\mu
}+\frac{\alpha
_{4}}{2}gf^{abc}\overline{c}^{b}c^{c}+\frac{\alpha _{5}}{2}gf^{abi}\overline{%
c}^{b}c^{i}\right. \right.  \nonumber \\
&+&\left. \left. \frac{\alpha
_{6}}{2}gf^{abi}c^{b}\bar{c}^{i}\right)
+b^{i}\left( \partial ^{\mu }{A}_{\mu }^{i}+\frac{\alpha _{3}}{2}b^{i}+\frac{%
\alpha _{6}}{2}gf^{abi}c^{a}\bar{c}^{b}\right) +\bar{c}^{a}\partial
^{2}c^{a}+\bar{c}^{i}\partial ^{2}c^{i}\right.  \nonumber \\
&+&\left. gf^{abc}\partial ^{\mu }\bar{c}^{a}c^{b}A_{\mu
}^{c}+gf^{abi}\partial ^{\mu }\bar{c}^{a}c^{b}A_{\mu
}^{i}-gf^{abi}\partial
^{\mu }\bar{c}^{a}c^{i}A_{\mu }^{b}-gf^{abi}c^{a}\partial ^{\mu }\bar{c}%
^{i}A_{\mu }^{b}\right.  \nonumber \\
&+&\left. \alpha _{2}gf^{abi}\bar{c}^{a}\partial ^{\mu
}{c}^{b}A_{\mu }^{i}+\alpha _{2}gf^{abi}\bar{c}^{a}\partial ^{\mu
}{c}^{i}A_{\mu }^{b}-\alpha
_{2}g^{2}f^{abi}f^{bcj}\bar{c}^{a}c^{c}A_{\mu }^{i}A^{j\mu
}\right.  \nonumber \\
&+&\left. \alpha _{2}g^{2}f^{abi}f^{bcd}\bar{c}^{a}c^{d}A_{\mu
}^{c}A^{i\mu }+\alpha _{2}g^{2}f^{abi}f^{bcj}\bar{c}^{a}c^{j}A_{\mu
}^{c}A^{i\mu }+\alpha _{2}g^{2}f^{abi}f^{cdi}\bar{c}^{a}c^{d}A_{\mu
}^{c}A^{b\mu }\right.
\nonumber \\
&+&\left. \frac{\alpha _{4}}{4}g^{2}f^{abc}f^{adi}\bar{c}^{b}\bar{c}%
^{c}c^{d}c^{i}+\frac{\alpha _{4}}{8}g^{2}f^{abc}f^{ade}\bar{c}^{b}\bar{c}%
^{c}c^{d}c^{e}+\frac{\alpha _{5}}{8}g^{2}f^{abi}f^{cdi}\bar{c}^{a}\bar{c}%
^{b}c^{c}c^{d}\right.  \nonumber \\
&+&\left. \frac{\alpha _{6}}{2}g^{2}f^{abi}f^{acj}\bar{c}^{b}\bar{c}%
^{i}c^{c}c^{j}+\frac{\alpha _{6}}{4}g^{2}f^{abi}f^{acd}\bar{c}^{b}\bar{c}%
^{i}c^{c}c^{d}\right] \;.  \label{gf2}
\end{eqnarray}
For calculational purposes it is useful to eliminate the Lagrange
multipliers. This is done by means of the equations of motion
\begin{eqnarray}
\frac{\delta \left( S_{\mathrm{YM}}+S_{\mathrm{GF}}\right) }{\delta
{b}^{a}} &=&\alpha _{1}{b}^{a}+\partial ^{\mu }{A}_{\mu }^{a}+\alpha
_{2}gf^{abi}A_{\mu }^{b}A^{i\mu }+\frac{\alpha _{4}}{2}gf^{abc}\overline{c}%
^{b}c^{c}+\frac{\alpha
_{5}}{2}gf^{abi}\overline{c}^{b}c^{i}+\frac{\alpha
_{6}}{2}gf^{abi}c^{b}\bar{c}^{i},  \nonumber \\
\frac{\delta \left( S_{\mathrm{YM}}+S_{GF}\right) }{\delta {b}^{i}} &=&{%
\alpha _{3}}b^{i}+\partial ^{\mu }{A}_{\mu }^{i}+\frac{\alpha _{6}}{2}%
gf^{abi}c^{a}\bar{c}^{b}\;\;.  \label{beqs}
\end{eqnarray}
Substituting the expressions (\ref{beqs}) into the expression
(\ref{gf2}), we find
\begin{eqnarray}
S_{\mathrm{GF}} &=&\int {d^{4}}x\left[ -\frac{1}{2\alpha _{1}}\partial ^{\mu }{A}%
_{\mu }^{a}\partial ^{\nu }{A}_{\nu }^{a}-\frac{1}{2\alpha
_{3}}\partial
^{\mu }{A}_{\mu }^{i}\partial ^{\nu }{A}_{\nu }^{i}-\frac{\alpha _{2}}{%
\alpha _{1}}gf^{abi}\partial ^{\mu }A_{\mu }^{a}A_{\nu }^{b}A^{i\nu
}\right.
\nonumber \\
&-&\left. \frac{\alpha _{2}^{2}}{2\alpha
_{1}}g^{2}f^{abi}f^{acj}A_{\mu
}^{b}A^{i\mu }A_{\nu }^{c}A^{j\nu }+\bar{c}^{a}\partial ^{2}c^{a}+\bar{c}%
^{i}\partial ^{2}c^{i}+\frac{\alpha _{4}}{2\alpha _{1}}gf^{abc}\bar{c}%
^{a}\partial ^{\mu }{c}^{b}A_{\mu }^{c}\right.  \nonumber \\
&+&\left. \left( \frac{\alpha _{4}}{2\alpha _{1}}+1\right)
gf^{abc}\partial
^{\mu }\bar{c}^{a}c^{b}A_{\mu }^{c}+\left( \frac{\alpha _{6}}{2\alpha _{3}}%
+1\right) gf^{abi}\partial ^{\mu }\bar{c}^{a}c^{b}A_{\mu }^{i}+\left( \frac{%
\alpha _{6}}{2\alpha _{1}}+1\right) gf^{abi}A_{\mu }^{a}c^{b}\partial ^{\mu }%
\bar{c}^{i}\right.  \nonumber \\
&-&\left. \left( \frac{\alpha _{5}}{2\alpha _{1}}+1\right)
gf^{abi}\partial
^{\mu }\bar{c}^{a}c^{i}A_{\mu }^{b}-\frac{\alpha _{6}}{2\alpha _{1}}%
gf^{abi}\partial ^{\mu }{c}^{a}\bar{c}^{i}A_{\mu }^{b}+\left(
\frac{\alpha
_{6}}{2\alpha _{3}}+\alpha _{2}\right) gf^{abi}\bar{c}^{a}\partial ^{\mu }{c}%
^{b}A_{\mu }^{i}\right.  \nonumber \\
&+&\left. \left( \alpha _{2}-\frac{\alpha _{5}}{2\alpha _{1}}\right) gf^{abi}%
\bar{c}^{a}\partial ^{\mu }{c}^{i}A_{\mu }^{b}+\alpha
_{2}g^{2}\left(
f^{abi}f^{bcd}+\frac{\alpha _{4}}{2\alpha _{1}}f^{abd}f^{bci}\right) \bar{c}%
^{a}c^{d}A_{\mu }^{c}A^{i\mu }\right.  \nonumber \\
&+&\left. \left( \frac{\alpha _{5}}{2\alpha _{1}}+1\right) \alpha
_{2}g^{2}f^{abi}f^{bcj}\bar{c}^{a}c^{j}A_{\mu }^{c}A^{i\mu
}-\frac{\alpha _{2}\alpha _{6}}{2\alpha
_{1}}g^{2}f^{abi}f^{acj}c^{b}\bar{c}^{i}A_{\mu
}^{c}A^{j\mu }\right.  \nonumber \\
&-&\left. \alpha _{2}g^{2}f^{abi}f^{bcj}\bar{c}^{a}c^{c}A_{\mu
}^{i}A^{j\mu
}+\alpha _{2}g^{2}f^{abi}f^{cdi}\bar{c}^{a}c^{d}A_{\mu }^{c}A^{b\mu }+\frac{%
\alpha _{4}}{4}g^{2}\left( f^{abc}f^{adi}\right. \right.  \nonumber \\
&+&\left. \left. \frac{\alpha _{5}}{\alpha _{1}}f^{abd}f^{aci}\right) \bar{c}%
^{b}\bar{c}^{c}c^{d}c^{i}+\frac{g^{2}}{8}\left( \alpha
_{5}f^{abi}f^{cdi}+\alpha _{4}f^{abe}f^{cde}+\frac{\alpha
_{6}^{2}}{\alpha
_{3}}f^{aci}f^{bdi}\right. \right.  \nonumber \\
&+&\left. \left. \frac{\alpha _{4}^{2}}{\alpha
_{1}}f^{ace}f^{bde}\right)
\bar{c}^{a}\bar{c}^{b}c^{c}c^{d}+\frac{\alpha _{6}}{2}\left(
1+\frac{\alpha
_{5}}{2\alpha _{1}}\right) g^{2}f^{abi}f^{acj}\bar{c}^{b}\bar{c}%
^{i}c^{c}c^{j}+\frac{\alpha _{6}}{4}g^{2}\left(
f^{abi}f^{acd}\right. \right.
\nonumber \\
&-&\left. \left. \frac{\alpha _{4}}{\alpha _{1}}f^{abc}f^{adi}\right) \bar{c}%
^{b}\bar{c}^{i}c^{c}c^{d}\;\right] \;.  \label{gf3}
\end{eqnarray}

\sect{Interpolating among various gauges.} Let us now proceed by
showing how expression (\ref{gf3}) can be used to interpolate among
various gauges. We begin with the linear covariant gauges, including
the Landau gauge.
\subsection{Linear covariant and Landau gauge.}
The linear covariant gauges (LCG) correspond to the following gauge
fixing term
\begin{equation}
S_{\mathrm{LCG}}=s\int {d^{4}x}\;\bar{c}^{A}\left( \frac{\alpha
}{2}b^{A}+\partial ^{\mu }{A}_{\mu }^{A}\right) \;,  \label{lcg0}
\end{equation}
which yields
\begin{equation}
S_{\mathrm{LCG}}=\int {d^{4}x}\left[ \bar{c}^{A}\partial ^{\mu
}{D}_{\mu }^{AB}c^{B}+b^{A}\left( \frac{\alpha }{2}b^{A}+\partial
^{\mu }{A}_{\mu }^{A}\right) \right] \;.  \label{lcg1}
\end{equation}
Eliminating the Lagrange multiplier
\begin{equation}
S_{\mathrm{LCG}}=\int {d^{4}x}\left( \bar{c}^{A}\partial ^{\mu }{D}_{\mu }^{AB}c^{B}-%
\frac{1}{2\alpha }\partial ^{\mu }{A}_{\mu }^{A}\partial ^{\nu
}{A}_{\nu }^{A}\right) \;.  \label{lcg2}
\end{equation}
As usual, the limit $\alpha \rightarrow 0$, defines the Landau
gauge. In order to recover the linear covariant gauges from
expression (\ref{gf3}), the values of the gauge parameters should be
taken as is shown in Table 1, namely
\begin{equation}
\alpha _{2}=\alpha _{4}=\alpha _{5}=\alpha _{6}=0\;,  \label{l1}
\end{equation}
and
\begin{equation}
\alpha _{1}=\alpha _{3}=\alpha \;.  \label{l2}
\end{equation}
\begin{table}[t]
\centering
\begin{tabular}{|c|c|c|c|c|c|c|}
\hline
gauge parameter & $\alpha_1$ & $\alpha_2$ & $\alpha_3$ & $\alpha_4$ & $%
\alpha_5$ & $\alpha_6$ \\ \hline limit & $\alpha$ & $0$ & $\alpha$ &
$\alpha_6$ & $\alpha_6$ & $0$ \\ \hline
\end{tabular}
\caption{Values of the gauge parameters for the linear covariant
gauges.} \label{table1}
\end{table}
After this, expression (\ref{gf3}) yields
\begin{eqnarray}
S_{\mathrm{GF}} &=&\int d^{4}x\left( -\frac{1}{2\alpha }\partial
^{\mu }{A}_{\mu
}^{a}\partial ^{\nu }{A}_{\nu }^{a}-\frac{1}{2\alpha }\partial ^{\mu }{A}%
_{\mu }^{i}\partial ^{\nu }{A}_{\nu }^{i}+\bar{c}^{a}\partial ^{2}c^{a}+\bar{%
c}^{i}\partial ^{2}c^{i}\right.  \nonumber \\
&+&\left. gf^{abc}\partial ^{\mu }\bar{c}^{a}c^{b}A_{\mu
}^{c}+gf^{abi}\partial ^{\mu }\bar{c}^{a}c^{b}A_{\mu
}^{i}+gf^{abi}A_{\mu
}^{a}c^{b}\partial ^{\mu }\bar{c}^{i}-gf^{abi}\partial ^{\mu }\bar{c}%
^{a}c^{i}A_{\mu }^{b}\right) \;.  \label{lcg3}
\end{eqnarray}
Now, collecting together both diagonal and off-diagonal indices, we
can rewrite the above expression as
\begin{equation}
S_{\mathrm{GF}}=\int {d^{4}}x\left( -\frac{1}{2\alpha }\partial
^{\mu }{A}_{\mu }^{A}\partial ^{\nu }{A}_{\nu
}^{A}+\bar{c}^{A}\partial ^{2}c^{A}+gf^{ABC}\partial ^{\mu
}\bar{c}^{A}c^{B}A_{\mu }^{C}\right) \;. \label{lcg4}
\end{equation}
Upon integration by parts of the last term, we obtain
\begin{equation}
S_{\mathrm{GF}}=\int {d^{4}}x\left( -\frac{1}{2\alpha }\partial
^{\mu }{A}_{\mu }^{A}\partial ^{\nu }{A}_{\nu
}^{A}+\bar{c}^{A}\partial ^{\mu }{D}_{\mu }^{AB}c^{B}\right) \;,
\label{lcg5}
\end{equation}
which is precisely expression (\ref{lcg2}). In particular, the
Landau gauge is thus recovered in the limit $\alpha \rightarrow 0$.
\subsection{Curci-Ferrari gauge.}
Let us consider now the Curci-Ferrari gauge. It is a nonlinear
covariant gauge, introduced in
\cite{A14,Delbourgo:1981cm,Baulieu:1981sb}, specified by
\begin{equation}
S_{\mathrm{CF}}=s\int {d^{4}x}\left( \bar{c}^{A}\left( \frac{\alpha }{2}%
b^{A}+\partial ^{\mu }{A}_{\mu }^{A}\right) -\frac{\alpha }{4}gf^{ABC}\bar{c}%
^{B}c^{C}\right) \;,  \label{cf0}
\end{equation}
namely
\begin{equation}
S_{\mathrm{CF}}=\int \! \! {d^{4}x}\left( \! b^{A}\left(
\frac{\alpha }{2}b^{A}+\partial ^{\mu
}{A}_{\mu }^{A}-\frac{\alpha }{2}gf^{ABC}\bar{c}^{B}c^{C}\right) +\bar{c}%
^{A}\partial ^{\mu }{D}_{\mu }^{AB}c^{B}-\frac{\alpha }{8}g^{2}f^{ABC}f^{CDE}%
\bar{c}^{A}\bar{c}^{B}c^{D}c^{E}\right) ,  \label{cf1}
\end{equation}
which, after elimination of the Lagrange multiplier, becomes
\begin{eqnarray}
S_{\mathrm{CF}} &=&\int {d^{4}x}\left( -\frac{1}{2\alpha }\partial
^{\mu }{A}_{\mu }^{A}\partial ^{\nu }{A}_{\nu
}^{A}+\bar{c}^{A}\partial ^{\mu }{D}_{\mu
}^{AB}c^{B}+\frac{g}{2}gf^{ABC}\partial ^{\mu }{A}_{\mu }^{A}\bar{c}%
^{B}c^{C}\right.  \nonumber \\
&+&\left. \frac{\alpha }{8}g^{2}f^{AEC}f^{CBD}\bar{c}^{A}\bar{c}%
^{B}c^{D}c^{E}\right) \;,  \label{cf2}
\end{eqnarray}
where use has been made of the Jacobi identity (\ref{jacobi}). To
obtain the Curci-Ferrari gauge from expression (\ref{gf3}) the gauge
parameters have to be taken as in Table 2, i.e.
\begin{eqnarray}
\alpha _{2} &=&0\;,  \label{cfb} \\
\alpha _{1} &=&\alpha _{3}=-\alpha _{4}=-\alpha _{5}=-\alpha
_{6}=\alpha \;, \nonumber
\end{eqnarray}
\begin{table}[t]
\centering
\begin{tabular}{|c|c|c|c|c|c|c|}
\hline
gauge parameter & $\alpha_1$ & $\alpha_2$ & $\alpha_3$ & $\alpha_4$ & $%
\alpha_5$ & $\alpha_6$ \\ \hline limit & $-\alpha_6$ & $0$ &
$-\alpha_6$ & $\alpha_6$ & $\alpha_6$ & $-\alpha$
\\ \hline
\end{tabular}
\caption{Values of the gauge parameters for the Curci-Ferrari
gauge.} \label{table2}
\end{table}
This leads to
\begin{eqnarray}
S_{\mathrm{CF}} &=&\int {d^{4}}x\left( -\frac{1}{2\alpha }\partial
^{\mu }{A}_{\mu
}^{a}\partial ^{\nu }{A}_{\nu }^{a}-\frac{1}{2\alpha }\partial ^{\mu }{A}%
_{\mu }^{i}\partial ^{\nu }{A}_{\nu }^{i}+\bar{c}^{a}\partial ^{2}c^{a}+\bar{%
c}^{i}\partial ^{2}c^{i}-\frac{1}{2}gf^{abc}\bar{c}^{a}\partial ^{\mu }{c}%
^{b}A_{\mu }^{c}\right.  \nonumber \\
&+&\left. \frac{1}{2}gf^{abc}\partial ^{\mu }\bar{c}^{a}c^{b}A_{\mu }^{c}+%
\frac{1}{2}gf^{abi}\partial ^{\mu }\bar{c}^{a}c^{b}A_{\mu }^{i}+\frac{1}{2}%
gf^{abi}A_{\mu }^{a}c^{b}\partial ^{\mu }\bar{c}^{i}-\frac{1}{2}%
gf^{abi}\partial ^{\mu }\bar{c}^{a}c^{i}A_{\mu }^{b}\right.  \nonumber \\
&+&\left. \frac{1}{2}gf^{abi}\partial ^{\mu }{c}^{a}\bar{c}^{i}A_{\mu }^{b}-%
\frac{1}{2}gf^{abi}\bar{c}^{a}\partial ^{\mu }{c}^{b}A_{\mu }^{i}+\frac{1}{2}%
gf^{abi}\bar{c}^{a}\partial ^{\mu }{c}^{i}A_{\mu }^{b}\right.  \nonumber \\
&-&\left. \frac{\alpha }{4}g^{2}\left(
f^{abc}f^{adi}-f^{abd}f^{aci}\right)
\bar{c}^{b}\bar{c}^{c}c^{d}c^{i}-\alpha \frac{g^{2}}{8}\left(
f^{abi}f^{cdi}+f^{abe}f^{cde}-f^{aci}f^{bdi}\right. \right.  \nonumber \\
&-&\left. \left. f^{ace}f^{bde}\right) \bar{c}^{a}\bar{c}^{b}c^{c}c^{d}-%
\frac{\alpha }{4}g^{2}f^{abi}f^{acj}\bar{c}^{b}\bar{c}^{i}c^{c}c^{j}-\frac{%
\alpha }{4}g^{2}\left( f^{abi}f^{acd}+f^{abc}f^{adi}\right) \bar{c}^{b}\bar{c%
}^{i}c^{c}c^{d}\right) \;.  \nonumber \\
&&  \label{cf3}
\end{eqnarray}
Collecting the various indices and using the Jacobi identities
(\ref{jacobi2}) for the quartic ghost interaction terms, we obtain
\begin{eqnarray}
S_{\mathrm{CF}} &=&\int {d^{4}}x\left( -\frac{1}{2\alpha }\partial
^{\mu }{A}_{\mu
}^{A}\partial ^{\nu }{A}_{\nu }^{A}+\bar{c}^{A}\partial ^{2}c^{A}-\frac{1}{2}%
gf^{ABC}\bar{c}^{A}\partial ^{\mu }{c}^{B}A_{\mu }^{C}+\frac{1}{2}%
gf^{ABC}\partial ^{\mu }\bar{c}^{A}c^{B}A_{\mu }^{C}\right.  \nonumber \\
&+&\left. \frac{\alpha }{4}g^{2}f^{abi}f^{acd}\bar{c}^{b}\bar{c}%
^{c}c^{d}c^{i}+\alpha \frac{g^{2}}{8}\left(
f^{ade}f^{ebc}+f^{adi}f^{ibc}\right) \bar{c}^{a}\bar{c}^{b}c^{c}c^{d}-\frac{%
\alpha
}{4}g^{2}f^{abi}f^{acj}\bar{c}^{b}\bar{c}^{i}c^{c}c^{j}\right.
\nonumber \\
&-&\left. \frac{\alpha }{4}g^{2}f^{abd}f^{aci}\bar{c}^{b}\bar{c}%
^{i}c^{c}c^{d}\right) \;.  \label{cf4}
\end{eqnarray}
In other words
\begin{eqnarray}
S_{\mathrm{CF}} &=&\int {d^{4}}x\left( -\frac{1}{2\alpha }\partial
^{\mu }{A}_{\mu
}^{A}\partial ^{\nu }{A}_{\nu }^{A}+\bar{c}^{A}\partial ^{2}c^{A}-\frac{1}{2}%
gf^{ABC}\bar{c}^{A}\partial ^{\mu }{c}^{B}A_{\mu }^{C}+ \frac{\alpha }{8}g^{2}f^{EAD}f^{EBC}\bar{c}^{A}\bar{c}%
^{B}c^{C}c^{D}\right.  \nonumber \\
&+&\left.\frac{1}{2} gf^{ABC}\partial ^{\mu }\bar{c}^{A}c^{B}A_{\mu
}^{C}\right) \;. \label{cf5}
\end{eqnarray}
Finally, integrating the last term of the above expression by parts,
eq.(\ref {cf5}), yields
\begin{eqnarray}
S_{\mathrm{CF}} &=&\int {d^{4}}x\left( -\frac{1}{2\alpha }\partial
^{\mu }{A}_{\mu }^{A}\partial ^{\nu }{A}_{\nu
}^{A}+\bar{c}^{A}\partial ^{\mu }{D}_{\mu
}^{AB}c^{B}+\frac{1}{2}gf^{ABC}\bar{c}^{A}c^{B}\partial ^{\mu
}{A}_{\mu
}^{C}\right.  \nonumber \\
&+&\left. \frac{\alpha }{8}g^{2}f^{EAD}f^{EBC}\bar{c}^{A}\bar{c}%
^{B}c^{C}c^{D}\right) \;,  \label{cf6}
\end{eqnarray}
which reproduces the Curci-Ferrari gauge fixing term, (\ref{cf2}).
\subsection{Maximal Abelian gauge.}
The so called MAG gauge condition amounts to fixing the value of the
covariant derivative, $D_{\mu }^{ab}A^{\mu b}$, of the off-diagonal
components by requiring that the functional
\begin{equation}
\mathcal{R}[A]=(VT)^{-1}\int d^{4}x\left( A_{\mu }^{a}A^{\mu
a}\right) \;, \label{MAGfunctional}
\end{equation}
attains a minimum with respect to the local gauge transformations.
This corresponds to imposing
\begin{equation}
D_{\mu }^{ab}A^{\mu b}=0\;.  \label{MAGdiff}
\end{equation}
However, this condition being non-linear, a quartic ghost
self-interaction term is required for renormalizability purposes.
The corresponding gauge fixing term turns out to be
\cite{Min:bx,Fazio:2001rm}
\begin{equation}
S_{\mathrm{MAG}}^{\mathrm{off}}=s\int d^{4}x\left[
\,\overline{c}^{a}\left( D_{\mu
}^{ab}A^{b\mu }+\frac{\alpha }{2}b^{a}\right) -\frac{\alpha }{2}gf\,^{abi}%
\overline{c}^{a}\overline{c}^{b}c^{i}-\frac{\alpha }{4}gf\,^{abc}c^{a}%
\overline{c}^{b}\overline{c}^{c}\right] \;,  \label{smn}
\end{equation}
and the real MAG, eq.(\ref{MAGdiff}), is obtained by setting
$\alpha=0$ after renormalization. The MAG condition allows for a
residual local $U(1)^{N-1}$ invariance with respect to the diagonal
subgroup of $SU(N)$. In order to have a complete quantization of the
theory, one has to fix this Abelian gauge freedom by
means of a suitable further gauge condition on the diagonal components $%
A_{\mu }^{i}$ of the gauge field. A common choice for the Abelian
gauge fixing, also adopted in the lattice papers
\cite{Amemiya:1998jz,Bornyakov:2003ee}, is the Landau gauge, given
by
\begin{equation}
S_{\mathrm{MAG}}^{\mathrm{diag}}=s\int
d^{4}x\overline{c}^{i}\partial _{\mu }A^{i\mu }\;. \label{abgf}
\end{equation}
The complete gauge fixing is, thus, the sum of expressions (\ref{smn}) and (%
\ref{abgf}), namely
\begin{equation}
S_{\mathrm{MAG}}=s\int {d^{4}}x\left[\overline{c}^{a}\left( D_{\mu
}^{ab}A^{b\mu
}+\frac{\alpha }{2}b^{a}\right) +\overline{c}^{i}\partial _{\mu }{A}^{i\mu }-%
\frac{\alpha
}{2}gf^{abi}\overline{c}^{a}\overline{c}^{b}c^{i}-\frac{\alpha
}{4}gf^{abc}c^{a}\overline{c}^{b}\overline{c}^{c}\right] \;.
\label{mag1}
\end{equation}
Working out expression (\ref{mag1}) we find
\begin{eqnarray}
S_{\mathrm{MAG}} &=&\int {d^{4}}x\left[ b^{a}\left( \partial ^{\mu
}{A}_{\mu
}^{a}-gf^{abi}A_{\mu }^{b}A^{i\mu }+\frac{\alpha }{2}b^{a}-\alpha {g}f^{abi}%
\bar{c}^{b}c^{i}-\frac{\alpha }{2}gf^{abc}\bar{c}^{b}c^{c}\right)
+b^{i}\partial ^{\mu }{A}_{\mu }^{i}\right.  \nonumber \\
&+&\left. \bar{c}^{a}D_{\mu }^{ab}D^{bc\mu
}c^{c}+\bar{c}^{i}\partial
^{2}c^{i}+gf^{abi}A_{\mu }^{a}\partial ^{\mu }\bar{c}^{b}c^{i}+gf^{abi}A_{%
\mu }^{a}\bar{c}^{b}\partial ^{\mu }{c}^{i}+gf^{abi}A_{\mu
}^{a}c^{b}\partial ^{\mu }\bar{c}^{i}\right.  \nonumber \\
&+&\left.gf^{abc}\partial ^{\mu }\bar{c}%
^{a}c^{b}A_{\mu }^{c}- g^{2}f^{abi}f^{bcj}\bar{c}^{a}c^{i}A_{\mu
}^{c}A^{j\mu }-g^{2}f^{abi}f^{bcd}\bar{c}^{a}c^{d}A_{\mu
}^{c}A^{i\mu }\right. \nonumber
\\
&-&\left.g^{2}f^{abi}f^{cdi}\bar{c}^{a}c^{d}A_{\mu
}^{b}A^{c\mu }- \frac{\alpha }{4}g^{2}f^{abi}f^{cdi}\bar{c}^{a}\bar{c}%
^{b}c^{c}c^{d}-\frac{\alpha }{4}g^{2}f^{abc}f^{adi}\bar{c}^{b}\bar{c}%
^{c}c^{d}c^{i}\right.\nonumber\\&-&\left.\frac{\alpha }{8}g^{2}f^{abc}f^{ade}\bar{c}^{b}\bar{c}%
^{c}c^{d}c^{e}\right] \;.  \label{mag2}
\end{eqnarray}
The task of eliminating the off-diagonal Lagrange multipliers is
straightforward. To eliminate the diagonal Lagrange multiplier we
have to introduce a parameter $\beta $ via the term
$\frac{\beta}{2}b^i b^i$, and then let $\beta \rightarrow 0$ to
recover the transversality of the diagonal gauge field. This will
lead to the following gauge fixing
\begin{eqnarray}
S_{\mathrm{MAG}} &=&\lim_{\beta \rightarrow 0}\int {d^{4}}x\,\left[
-\frac{1}{2\alpha
}\partial ^{\mu }{A}_{\mu }^{a}\partial ^{\nu }{A}_{\nu }^{a}-\frac{1}{%
2\beta }\partial ^{\mu }{A}_{\mu }^{i}\partial ^{\nu }{A}_{\nu }^{i}+\frac{g%
}{\alpha }f^{abi}\partial ^{\mu }{A}_{\mu }^{a}A_{\nu }^{a}A^{i\nu
}\right.
\nonumber \\
&-&\!\!\left. \frac{g^{2}}{2\alpha }f^{abi}f^{acj}A_{\mu
}^{b}A^{i\mu }A_{\nu }^{c}A^{j\nu }+\bar{c}^{a}D_{\mu }^{ab}D^{bc\mu
}c^{c}+\bar{c}^{i}\partial
^{2}c^{i}+gf^{abi}A_{\mu }^{a}c^{b}\partial ^{\mu }\bar{c}^{i}+\frac{1}{2}%
gf^{abc}\partial ^{\mu }\bar{c}^{a}c^{b}A_{\mu }^{c}\right.  \nonumber \\
&-&\left. \frac{1}{2}gf^{abc}\bar{c}^{a}\partial ^{\mu }{c}^{b}A_{\mu }^{c}-%
\frac{g^{2}}{2}\left( 2f^{abi}f^{bcd}-f^{abd}f^{bci}\right) \bar{c}%
^{a}c^{d}A_{\mu }^{c}A^{i\mu
}-g^{2}f^{abi}f^{cdi}\bar{c}^{a}c^{d}A_{\mu
}^{b}A^{c\mu }\right.  \nonumber \\
&-&\left. \frac{\alpha }{4}g^{2}\left(
2f^{abi}f^{acd}+f^{abc}f^{adi}\right)
\bar{c}^{b}\bar{c}^{c}c^{d}c^{i}\right.\nonumber\\&-&\left.\frac{\alpha
}{8}g^{2}\left(
2f^{bci}f^{dei}+f^{abc}f^{ade}-f^{abd}f^{ace}\right) \bar{c}^{b}\bar{c}%
^{c}c^{d}c^{e}\vphantom{-\frac{1}{2\alpha }\partial ^{\mu }{A}_{\mu
}^{a}\partial ^{\nu }{A}_{\nu }^{a}}\right] \;.    \label{mag3}
\end{eqnarray}
Expression (\ref{mag3}) is recovered from the interpolating gauge
fixing (\ref{gf3}), once the gauge parameters are taken as in Table
3,
\begin{eqnarray}
\alpha _{1} &=&-\,\alpha _{4}=-\,\frac{\alpha _{5}}{2}=\alpha \;,
\label{mag3b}
\\
\alpha _{2} &=&-\,1\;,\;\alpha _{6}=0\;,  \nonumber
\end{eqnarray}
and then
\begin{equation}
\alpha _{3}=\beta \rightarrow 0\;.  \label{mag3c}
\end{equation}
\begin{table}[t]
\centering
\begin{tabular}{|c|c|c|c|c|c|c|}
\hline
gauge parameter & $\alpha_1$ & $\alpha_2$ & $\alpha_3$ & $\alpha_4$ & $%
\alpha_5$ & $\alpha_6$ \\ \hline
limit & $\alpha$ & $-1$ & $\beta\stackrel{end}{\rightarrow}0$ & $-\alpha$ & $%
-2\alpha$ & $0$ \\ \hline
\end{tabular}
\caption{Values of the gauge parameters for the maximal Abelian
gauge.} \label{table3}
\end{table}
Therefore, expression (\ref{gf3}) becomes
\begin{eqnarray}
S_{\mathrm{GF}} &=&\lim_{\beta \rightarrow 0}\int {d^{4}}x\left[ -\frac{1}{2\alpha }%
\partial ^{\mu }{A}_{\mu }^{a}\partial ^{\nu }{A}_{\nu }^{a}-\frac{1}{2\beta
}\partial ^{\mu }{A}_{\mu }^{i}\partial ^{\nu }{A}_{\nu }^{i}+\frac{1}{%
\alpha }gf^{abi}\partial ^{\mu }A_{\mu }^{a}A_{\nu }^{b}A^{i\nu
}\right.
\nonumber \\
&-&\left. \frac{1}{2\alpha }g^{2}f^{abi}f^{acj}A_{\mu }^{b}A^{i\mu
}A_{\nu
}^{c}A^{j\nu }+\bar{c}^{a}\partial ^{2}c^{a}+\bar{c}^{i}\partial ^{2}c^{i}-%
\frac{1}{2}gf^{abc}\bar{c}^{a}\partial ^{\mu }{c}^{b}A_{\mu
}^{c}\right.
\nonumber \\
&+&\left. \frac{1}{2}gf^{abc}\partial ^{\mu }\bar{c}^{a}c^{b}A_{\mu
}^{c}+gf^{abi}\partial ^{\mu }\bar{c}^{a}c^{b}A_{\mu
}^{i}+gf^{abi}A_{\mu
}^{a}c^{b}\partial ^{\mu }\bar{c}^{i}-gf^{abi}\bar{c}^{a}\partial ^{\mu }{c}%
^{b}A_{\mu }^{i}\right.  \nonumber \\
&-&\left. g^{2}\left( f^{abi}f^{bcd}-\frac{1}{2}f^{abd}f^{bci}\right) \bar{c}%
^{a}c^{d}A_{\mu }^{c}A^{i\mu
}+g^{2}f^{abi}f^{bcj}\bar{c}^{a}c^{c}A_{\mu }^{i}A^{j\mu
}-g^{2}f^{abi}f^{cdi}\bar{c}^{a}c^{d}A_{\mu }^{c}A^{b\mu
}\right.  \nonumber \\
&-&\left. \frac{\alpha }{4}g^{2}\left(
f^{abc}f^{adi}-2f^{abd}f^{aci}\right)
\bar{c}^{b}\bar{c}^{c}c^{d}c^{i}\right.\nonumber\\&-&\left.\alpha
\frac{g^{2}}{8}\left(
2f^{abi}f^{cdi}+f^{abe}f^{cde}-f^{ace}f^{bde}\right) \bar{c}^{a}\bar{c}%
^{b}c^{c}c^{d}\right] \;,  \label{mag4}
\end{eqnarray}
which, after rearranging some indices, yields the expression
(\ref{mag3}).
\subsection{Modified maximal Abelian gauge.}
For completeness, let us include in the present analysis also a
slightly modified version of the maximal Abelian gauge. This gauge,
discussed in \cite{Dudal:2002ye}, is obtained by making the
following choice for the diagonal gauge fixing term
\begin{equation}
S_{\mathrm{MAG}}^{\mathrm{diag}}=s\int {d^{4}x}\left(
\bar{c}^{i}\partial ^{\mu }{A}_{\mu }^{i}+gf^{abi}\bar{c}^{a}A_{\mu
}^{b}A^{i\mu }\right) \;, \label{mmag0}
\end{equation}
Thus, for the complete gauge gauge fixing in the modified maximal
Abelian gauge, MMAG, we have
\begin{equation}
S_{\mathrm{MMAG}}=s\,\int {d^{4}}x\,\left( \overline{c}^{a}\left( \partial ^{\mu }{A}%
_{\mu }^{a}+\frac{\alpha }{2}b^{a}\right) +\overline{c}^{i}\partial _{\mu }{A%
}^{i\mu }-\frac{\alpha }{2}gf^{abi}\overline{c}^{a}\overline{c}^{b}c^{i}-%
\frac{\alpha
}{4}gf^{abc}c^{a}\overline{c}^{b}\overline{c}^{c}\right) \;,
\label{mmag1}
\end{equation}
namely
\begin{eqnarray}
S_{\mathrm{MMAG}} &=&\int {d^{4}}x\left[ b^{a}\left( \partial ^{\mu }{A}_{\mu }^{a}+%
\frac{\alpha }{2}b^{a}-\alpha {g}f^{abi}\bar{c}^{b}c^{i}-\frac{\alpha }{2}%
gf^{abc}\bar{c}^{b}c^{c}\right) +b^{i}\partial _{\mu }{A}^{i\mu }+\bar{c}%
^{a}\partial ^{\mu }{D}_{\mu }^{ab}c^{b}\right.  \nonumber \\
&+&\left. \bar{c}^{i}\partial ^{2}c^{i}+gf^{abi}A_{\mu
}^{a}c^{b}\partial ^{\mu }\bar{c}^{i}-gf^{abc}\partial ^{\mu
}\bar{c}^{a}A_{\mu }^{b}c^{c}-gf^{abi}\partial ^{\mu
}\bar{c}^{a}A_{\mu }^{b}c^{i}-\frac{\alpha
}{4}g^{2}f^{abc}f^{adi}\bar{c}^{b}\bar{c}^{c}c^{d}c^{i}\right.  \nonumber \\
&-&\left. \frac{\alpha }{4}g^{2}f^{abi}f^{cdi}\bar{c}^{a}\bar{c}%
^{b}c^{c}c^{d}-\frac{\alpha }{8}g^{2}f^{abe}f^{cde}\bar{c}^{a}\bar{c}%
^{b}c^{c}c^{d}\right] \;,  \label{mmag2}
\end{eqnarray}
which, upon elimination of the Lagrange multipliers, becomes
\begin{eqnarray}
S_{\mathrm{MMAG}} &=&\lim_{\beta \rightarrow 0}\int {d^{4}}x\left[ -\frac{1}{%
2\alpha }\partial ^{\mu }{A}_{\mu }^{a}\partial ^{\nu }{A}_{\nu }^{a}-\frac{1%
}{2\beta }\partial _{\mu }{A}^{i\mu }\partial _{\nu }{A}^{i\nu }+\bar{c}%
^{a}\partial ^{\mu }{D}_{\mu }^{ab}c^{b}+\bar{c}^{i}\partial
^{2}c^{i}\right.
\nonumber \\
&+&\left. gf^{abi}A_{\mu }^{a}c^{b}\partial ^{\mu }\bar{c}^{i}-\frac{g}{2}%
f^{abc}\bar{c}^{a}\partial ^{\mu }{c}^{b}A_{\mu }^{c}+\frac{g}{2}%
f^{abc}\partial ^{\mu }\bar{c}^{a}c^{b}A_{\mu }^{c}+gf^{abi}\bar{c}%
^{a}A_{\mu }^{b}\partial ^{\mu }{c}^{i}\right.  \nonumber \\
&-&\left. \frac{\alpha }{4}g^{2}\left(
2f^{abi}f^{acd}+f^{abc}f^{adi}\right)
\bar{c}^{b}\bar{c}^{c}c^{d}c^{i}\right.\nonumber\\&-&\left.\frac{\alpha
}{8}g^{2}\left(
2f^{abi}f^{cdi}+f^{abe}f^{cde}+f^{ade}f^{bce}\right) \bar{c}^{a}\bar{c}%
^{b}c^{c}c^{d}\vphantom{-\frac{1}{2\alpha }\partial ^{\mu }{A}_{\mu
}^{a}\partial ^{\nu }{A}_{\nu }^{a}}\right] \;.   \label{mmag3}
\end{eqnarray}
\noindent Again, expression (\ref{mmag3}) follows from (\ref{gf3})
by setting the gauge parameters as in Table 4,
\begin{eqnarray}
\alpha _{1} &=&-\,\alpha _{4}=-\,\frac{\alpha _{5}}{2}=\alpha \;,
\label{mmaga}
\\
\alpha _{2} &=&\;\alpha _{6}=0\;,  \nonumber
\end{eqnarray}
and
\[
\alpha _{3}=\beta \rightarrow 0\;.
\]
Thus
\begin{table}[t]
\centering
\begin{tabular}{|c|c|c|c|c|c|c|}
\hline
gauge parameter & $\alpha_1$ & $\alpha_2$ & $\alpha_3$ & $\alpha_4$ & $%
\alpha_5$ & $\alpha_6$ \\ \hline
limit & $\alpha$ & $0$ & $\beta\stackrel{end}{\rightarrow}0$ & $-\alpha$ & $%
-2\alpha$ & $0$ \\ \hline
\end{tabular}
\caption{Values of the gauge parameters for the modified maximal
Abelian gauge.} \label{table4}
\end{table}
\begin{eqnarray}
S_{\mathrm{GF}} &=&\lim_{\beta \rightarrow 0}\int {d^{4}}x\left( -\frac{1}{2\alpha }%
\partial ^{\mu }{A}_{\mu }^{a}\partial ^{\nu }{A}_{\nu }^{a}-\frac{1}{2\beta
}\partial ^{\mu }{A}_{\mu }^{i}\partial ^{\nu }{A}_{\nu }^{i}+\bar{c}%
^{a}\partial ^{2}c^{a}+\bar{c}^{i}\partial ^{2}c^{i}-\frac{1}{2}gf^{abc}\bar{%
c}^{a}\partial ^{\mu }{c}^{b}A_{\mu }^{c}\right.   \nonumber \\
&+&\left. \frac{1}{2}gf^{abc}\partial ^{\mu }\bar{c}^{a}c^{b}A_{\mu
}^{c}+gf^{abi}\partial ^{\mu }\bar{c}^{a}c^{b}A_{\mu
}^{i}+gf^{abi}A_{\mu
}^{a}c^{b}\partial ^{\mu }\bar{c}^{i}+gf^{abi}\bar{c}^{a}\partial ^{\mu }{c}%
^{i}A_{\mu }^{b}\right.   \nonumber \\
&-&\left. \frac{\alpha }{4}g^{2}\left(
f^{abc}f^{adi}-2f^{abd}f^{aci}\right)
\bar{c}^{b}\bar{c}^{c}c^{d}c^{i}\right.\nonumber\\&-&\left.\alpha
\frac{g^{2}}{8}\left(
2f^{abi}f^{cdi}+f^{abe}f^{cde}-f^{ace}f^{bde}\right) \bar{c}^{a}\bar{c}%
^{b}c^{c}c^{d}\right) \;.  \nonumber \\  \label{mmag4}
\end{eqnarray}
By rearranging the indices of the quartic ghost terms by means of
the Jacobi identities (\ref{jacobi2}), it is then easy to show that
\begin{equation}
S_{\mathrm{GF}}=S_{\mathrm{MMAG}}\;.
\end{equation}

\sect{Multiplicative renormalizability of the interpolating gauge.}
To prove renormalizability, let us first write down the
Slavnov-Taylor identity corresponding to the BRST\ invariance of the
classical action. Following the algebraic renormalization setup
\cite{Piguet:1995er},  we introduce a set of external sources
coupled to the nonlinear BRST\ transformations $\left(
\ref{brst}\right) $ of the fields,
\begin{equation}
S_{\mathrm{ext}}=s\int {d^{4}x}\left( -\Omega ^{a\mu }A_{\mu
}^{a}-\Omega ^{i\mu }A_{\mu }^{i}+L^{a}c^{a}+L^{i}c^{i}\right) \;,
\label{ext}
\end{equation}
with
\begin{equation}
s\Omega _{\mu }^{a}=s\Omega _{\mu }^{i}=sL^{a}=sL^{i}=0\;,
\end{equation}
or
\begin{eqnarray}
S_{\mathrm{ext}} &=&\int {d^{4}x}\left[ -\Omega ^{a\mu }\left( D_{\mu }^{ab}{c}%
^{b}+gf^{abc}A_{\mu }^{b}c^{c}+gf^{abi}A_{\mu }^{b}c^{i}\right)
-\Omega ^{i\mu }\left( \partial _{\mu }{c}^{i}+gf^{abi}A_{\mu
}^{a}c^{b}\right)
\right.   \nonumber \\
&+&\left. L^{a}\left(
gf^{abi}c^{b}c^{i}+\frac{g}{2}f^{abc}c^{b}c^{c}\right)
+\frac{g}{2}f^{abi}c^{a}c^{b}L^{i}\right] \;.  \label{ext2}
\end{eqnarray}
Therefore, it follows that the complete action $\Sigma $
\begin{equation}
\Sigma =S_{YM}+S_{GF}+S_{ext}\;,  \label{complaction}
\end{equation}
obeys the Slavnov-Taylor identity
\begin{equation}
\mathcal{S}(\Sigma )=\int {d^{4}}x\left( \frac{\delta \Sigma
}{\delta \Omega
^{a\mu }}\frac{\delta \Sigma }{\delta A_{\mu }^{a}}+\frac{\delta \Sigma }{%
\delta \Omega ^{i\mu }}\frac{\delta \Sigma }{\delta A_{\mu }^{i}}+\frac{%
\delta \Sigma }{\delta L^{a}}\frac{\delta \Sigma }{\delta c^{a}}+\frac{%
\delta \Sigma }{\delta L^{i}}\frac{\delta \Sigma }{\delta c^{i}}+b^{a}\frac{%
\delta \Sigma }{\delta \overline{c}^{a}}+b^{i}\frac{\delta \Sigma
}{\delta \overline{c}^{i}}\right) =0\;.  \label{ste}
\end{equation}
In order to characterize the most general invariant counterterm
which can be freely added to any order of perturbation theory, we
perturb the classical action $\Sigma $ by adding an arbitrary
integrated local polynomial $\Sigma ^{\mathrm{count}}$ in the fields
and external sources of dimension bounded by four and with zero
ghost number, and we require that the perturbed action $\left(
\Sigma +\eta \Sigma ^{\mathrm{count}}\right) $ satisfies the same
Ward identities as $\Sigma $ to the first order in the perturbation
parameter $\eta$, being,
\begin{equation}
\mathcal{S}(\Sigma +\eta\Sigma ^{\mathrm{count}})=0\;+O(\eta
^{2})\;, \label{st1}
\end{equation}
This amounts to imposing the following conditions on $\Sigma ^{\mathrm{count}%
}$
\begin{equation}
\mathcal{S}_{\Sigma }\Sigma ^{\mathrm{count}}=0\;,  \label{bc}
\end{equation}
where $\mathcal{S}_{\Sigma }$ is the linearized Slavnov-Taylor
nilpotent operator,
\begin{eqnarray}
\mathcal{S}_{\Sigma } &=&\int d^{4}x\left( \frac{\delta \Sigma
}{\delta
\Omega ^{\mu a}}\frac{\delta }{\delta A_{\mu }^{a}}+\frac{\delta \Sigma }{%
\delta A_{\mu }^{a}}\frac{\delta }{\delta \Omega ^{\mu
a}}+\frac{\delta
\Sigma }{\delta \Omega ^{\mu i}}\frac{\delta }{\delta A_{\mu }^{i}}+\frac{%
\delta \Sigma }{\delta A_{\mu }^{i}}\frac{\delta }{\delta \Omega ^{\mu i}}+%
\frac{\delta \Sigma }{\delta L^{a}}\frac{\delta }{\delta
c^{a}}\right.
\nonumber \\
&+&\left. \frac{\delta \Sigma }{\delta c^{a}}\frac{\delta }{\delta L^{a}}+%
\frac{\delta \Sigma }{\delta L^{i}}\frac{\delta }{\delta
c^{i}}+\frac{\delta
\Sigma }{\delta c^{i}}\frac{\delta }{\delta L^{i}}+b^{a}\frac{\delta }{%
\delta \overline{c}^{a}}+b^{i}\frac{\delta }{\delta
\overline{c}^{i}}\right) \;,  \label{lb}
\end{eqnarray}
\begin{equation}
\mathcal{S}_{\Sigma }\mathcal{S}_{\Sigma }=0\;.  \label{st2}
\end{equation}
{}From the conditions $\left( \ref{bc}\right) $, it turns out that
the most general invariant counterterm $\Sigma ^{\mathrm{count}}$
can be written as
\begin{equation}
\Sigma ^{\mathrm{count}}=a_{0}S_{\mathrm{YM}}+\mathcal{S}_{\Sigma
}\Delta ^{-1}\;, \label{counter}
\end{equation}
where $\Delta ^{-1}$ is a local polynomial in the fields and
sources, with dimension four and ghost number, which reads
\begin{eqnarray}
\Delta ^{-1} &=&\int {d^{4}x}\left( a_{1}\frac{\alpha _{1}}{2}b^{a}\bar{c}%
^{a}+a_{2}\alpha _{2}gf^{abi}\bar{c}^{a}A_{\mu }^{b}A^{i\mu }+a_{3}\frac{%
\alpha _{3}}{2}b^{i}\bar{c}^{i}+a_{4}\frac{\alpha _{4}}{4}gf^{abc}\bar{c}^{a}%
\bar{c}^{b}c^{c}\right.   \nonumber \\
&+&a_{5}\frac{\alpha _{5}}{4}gf^{abi}\bar{c}^{a}\bar{c}^{b}c^{i}+a_{6}\frac{%
\alpha _{6}}{2}gf^{abi}c^{a}\bar{c}^{b}\bar{c}^{i}+a_{7}\partial ^{\mu }{A}%
_{\mu }^{a}\bar{c}^{a}+a_{8}\partial ^{\mu }{A}_{\mu }^{i}\bar{c}%
^{i}+a_{9}\Omega _{\mu }^{a}A^{a\mu }+a_{10}\Omega _{\mu
}^{i}A^{i\mu }
\nonumber \\
&+&\left. a_{11}L^{a}c^{a}+a_{12}L^{i}c^{i}\right) \;, \label{delta}
\end{eqnarray}
where $a_{i}\;|\;i\in \{0,1,\ldots ,12\}$  are independent
parameters.
\begin{table}[t]
\centering
\begin{tabular}{|c|c|c|c|c|c|c|}
\hline & $A$ & $c$ & $\bar{c}$ & $b$ & $\Omega$ & $L$ \\ \hline
dimension & 1 & 0 & 2 & 2 & 3 & 4 \\
ghost number & 0 & 1 & $-1$ & 0 & $-1$ & $-2$ \\ \hline
\end{tabular}
\caption{Dimensions and ghost numbers} \label{table6}
\end{table}
These parameters can be reabsorbed by means of a multiplicative
renormalization of the gauge coupling constant,  of the gauge
parameters, and of the fields and sources, according to
\begin{equation}
\Sigma (\Phi _{0},\phi _{0},J_{0},j_{0},\xi _{0})=\Sigma (\Phi ,\phi
,J,j,\xi )+\eta \Sigma ^{\mathrm{count}}(\Phi ,\phi ,J,j,\xi )\;,
\end{equation}
where
\begin{eqnarray}
\Phi _{0} &=&Z_{\Phi }^{1/2}\Phi \;,  \nonumber \\
\phi _{0} &=&\tilde{Z}_{\phi }^{1/2}\phi \;,  \nonumber \\
J_{0} &=&Z_{J}J\;,  \nonumber \\
j_{0} &=&\tilde{Z}_{j}j\;,  \nonumber \\
\xi _{0} &=&Z_{\xi }\xi \;,
\end{eqnarray}
with
\begin{eqnarray}
\Phi  & = &\left\{ A^{a},b^{a},c^{a},\overline{c}^{a}\right\} \;,
\nonumber
\\
\phi  & = &\left\{ A^{i},b^{i},c^{i},\overline{c}^{i}\right\} \;,
\nonumber
\\
J & = &\left\{ \Omega ^{a},L^{a}\right\} \;,  \nonumber \\
j & = &\left\{ \Omega ^{i},L^{i}\right\} \;,  \nonumber \\
\xi  & = &\left\{ g,\alpha _{1},\alpha _{2},\alpha _{3},\alpha
_{4},\alpha _{5},\alpha _{6}\right\} \;.
\end{eqnarray}
For the renormalization of the off-diagonal and  diagonal gluons,
and of the coupling constant $g$, we find, respectively,
\begin{eqnarray}
Z_{A}^{1/2} &=&1+\eta \left( \frac{a_{0}}{2}+a_{9}\right) \;,
\nonumber
\\
\tilde{Z}_{A}^{1/2} &=&1+\eta \left( \frac{a_{0}}{2}+a_{10}\right)
\;,
\nonumber \\
Z_{g} &=&1-\eta \frac{a_{0}}{2}\;,  \label{ren1}
\end{eqnarray}
while the Lagrange multipliers renormalize as follows
\begin{eqnarray}
Z_{b}^{1/2} &=&1+\eta \left( a_{7}-\frac{a_{0}}{2}\right) \;,
\nonumber
\\
\tilde{Z}_{b}^{1/2} &=&1+\eta \left( a_{8}-\frac{a_{0}}{2}\right)
\;. \label{ren2}
\end{eqnarray}
The renormalization factors of the gauge parameters are
\begin{eqnarray}
Z_{\alpha _{1}} &=&1+\eta \left( a_{0}+a_{1}-2a_{7}\right) \;,
\nonumber
\\
Z_{\alpha _{2}} &=&1+\eta \left( a_{2}-a_{7}\right) \;,  \nonumber \\
Z_{\alpha _{3}} &=&1+\eta \left( a_{0}+a_{3}-2a_{8}\right) \;,
\nonumber
\\
Z_{\alpha _{4}} &=&1+\eta \left( a_{0}+a_{4}-2a_{7}\right) \;,
\nonumber
\\
Z_{\alpha _{5}} &=&1+\eta \left( a_{0}+a_{5}-2a_{7}\right) \;,
\nonumber
\\
Z_{\alpha _{6}} &=&1+\eta \left( a_{0}+a_{6}-a_{7}-a_{8}\right) \;.
\label{ren3}
\end{eqnarray}
The renormalization of the ghost and anti-ghost fields are
constrained by the following self-consistent relations
\begin{eqnarray}
Z_{\bar{c}}^{1/2}Z_{c}^{1/2} &=&1+\eta \left( a_{7}-a_{11}\right)
\;,
\nonumber \\
\tilde{Z}_{\bar{c}}^{1/2}\tilde{Z}_{c}^{1/2} &=&1+\eta \left(
a_{8}-a_{12}\right) \;,  \nonumber \\
Z_{\bar{c}}^{1/2}\tilde{Z}_{c}^{1/2} &=&1+\eta \left(
a_{7}-a_{12}\right) \;,  \nonumber \\
\tilde{Z}_{\bar{c}}^{1/2}Z_{c}^{1/2} &=&1+\eta \left(
a_{8}-a_{11}\right) \;,  \label{reng}
\end{eqnarray}
which give
\begin{eqnarray}
Z_{c}^{1/2} &=&1+\eta \left( \frac{a_{7}}{2}+\frac{a_{8}}{2}-\frac{a_{11}%
}{2}+\frac{a_{12}}{2}\right) \;,  \nonumber \\
\tilde{Z}_{c}^{1/2} &=&1+\eta \left( \frac{a_{7}}{2}+\frac{a_{8}}{2}+%
\frac{a_{11}}{2}-\frac{a_{12}}{2}\right) \;,  \nonumber \\
Z_{\bar{c}}^{1/2} &=&1+\eta \left( \frac{a_{7}}{2}-\frac{a_{8}}{2}-\frac{%
a_{11}}{2}-\frac{a_{12}}{2}\right) \;,  \nonumber \\
\tilde{Z}_{\bar{c}}^{1/2} &=&1+\eta \left( -\frac{a_{7}}{2}+\frac{a_{8}}{%
2}-\frac{a_{11}}{2}-\frac{a_{12}}{2}\right) \;.  \label{ren4}
\end{eqnarray}
Finally, for the renormalization of the external sources
\begin{eqnarray}
Z_{\Omega } &=&1-\eta \left( \frac{a_{7}}{2}+\frac{a_{8}}{2}+a_{9}+\frac{%
a_{11}}{2}+\frac{a_{12}}{2}\right) \;,  \nonumber \\
\tilde{Z}_{\Omega } &=&1-\eta \left( \frac{a_{7}}{2}+\frac{a_{8}}{2}%
+a_{10}+\frac{a_{11}}{2}+\frac{a_{12}}{2}\right) \;,  \nonumber \\
Z_{L} &=&1+\eta \left( \frac{a_{0}}{2}-a_{7}-a_{8}-a_{12}\right) \;,
\nonumber \\
\tilde{Z}_{L} &=&1+\eta \left(
\frac{a_{0}}{2}-a_{7}-a_{8}-a_{11}\right) \;.  \label{ren5}
\end{eqnarray}
This completes the proof of the multiplicative renormalization of
the interpolating gauge fixing.

\sect{One loop renormalization.} We now record the details of the
explicit renormalization of the interpolating gauge at one loop
which requires the computation of the
renormalization constants $Z_{A}$, $\tilde{Z}_{A}$, $Z_{\psi }$, $Z_{g}$, $%
Z_{c}$, $Z_{\bar{c}}$, $\tilde{Z}_{c}$, $\tilde{Z}_{\bar{c}}$ and
$Z_{\alpha _{i}}$, with $i$ $=$ $1$, $\ldots$, $6$, and where
$Z_{\psi }$ is the quark wave function renormalization
constant\footnote{Although we did not consider matter fields in the
previous sections, it can be easily checked that the interpolating
gauge fixing term (\ref{gf3}) remains renormalizable if spinor
fields are included.}. Given the explicit form of the Lagrangian we
do not need to consider the renormalization of all three- and
four-point interactions. The procedure we adopt is to determine the
renormalization constants of the fields and of the gauge parameters
$\alpha_1,\alpha_3$ before deducing the explicit values of the
remaining four parameters $\alpha_2,\alpha_4, \alpha_5, \alpha_6$.
We use dimensional regularization
in $d$~$=$~$4$~$-$~$2\epsilon $ and subtract the divergences using the $%
\overline{\mbox{MS}}$ scheme. In previous work we carried out
similar renormalization in the Curci-Ferrari gauge and MAG,
\cite{A1,A2,A3,A4}, using the \textsc{Mincer} algorithm,
\cite{A5,A6}, written in \textsc{Form}, \cite{A7}, where the
diagrams for the appropriate Green's function were generated by the
\textsc{Qgraf} package, \cite{A8}. However, to determine the full
interpolating gauge renormalization constants it transpired that we
needed to renormalize several four-point interactions in addition to
various three-point functions. In this instance the \textsc{Mincer}
algorithm is not fully appropriate since it can only be applied to
two-point functions with massless propagators. For a four-point
function one would therefore need to nullify two external momenta
which could potentially introduce spurious infrared divergences. To
circumvent this difficulty for the four-point function
renormalization, we applied an alternative algorithm, still in
dimensional regularization, which involved nullifying all external
momenta but systematically introducing a temporary infrared mass
regularization according to the approach of \cite{A9,A10}. This
involved repeatedly replacing massless propagators by massive ones
via
\begin{equation}
\frac{1}{(k-p)^{2}}~=~\frac{1}{(k^{2}-m^{2})}~+~\frac{(2kp-p^{2}-m^{2})}{%
(k-p)^{2}(k^{2}-m^{2})}\;,
\end{equation}
until the contributions from the last term are finite and can be
neglected by Weinberg's theorem, \cite{A11}. The resulting one loop
vacuum bubble Feynman diagrams are then elementary to evaluate. For
the three-point functions we will still nullify an external momentum
but we will do so only in Green's functions where we know no
infrared singularities will be introduced. In either approach we
will still use the method of \cite{A12} to extract the explicit
renormalization constants. This involves computing the Green's
functions in terms of bare quantities before transforming to
renormalized variables and parameters. The form of the tree term
then allows one to fix the renormalization constants associated with
that Green's function. This approach avoids the need to consider
subtractions and moreover is more appropriate to performing
automatic symbolic manipulation computations. One additional point
arising from using the method of \cite {A12} is that we cannot
renormalize $\alpha _{5}$ itself since it never appears as an
isolated coefficient in one type of interaction. However, since it
appears in the combination $(1+\alpha _{5}/(2\alpha _{1}))$ we have
chosen to define a new parameter
\begin{equation}
\bar{\alpha}_{5}~=~1~+~\frac{\alpha _{5}}{2\alpha _{1}}\;,
\end{equation}
and determined its renormalization constant $Z_{\bar{\alpha}_{5}}$.
Clearly this choice does not invalidate the renormalizability of the
interpolating gauge. We note that to recover the various gauges
discussed in Sect. 2 the value of $\bar{\alpha}_{5}$ becomes $1$,
$\mbox{\small{$\frac{1}{2}$}}$, $0$ and $0$ respectively in the
linear covariant, Curci-Ferrari, maximal Abelian and modified
maximal Abelian gauges.

Since we are considering a gauge where the color group is split into
diagonal and off-diagonal sectors we briefly recall the essential
properties of the group algebra which is discussed in more depth in
\cite{A4}. In addition to the usual color group Casimirs $C_F$,
$C_A$ and $T_F$ for a general color group we define the dimensions
of the centre of the group as $N^d_{\!A}$ and the dimension of the
off-diagonal sector as $N^o_{\!A}$. Clearly
\begin{equation}
N^d_{\!A} ~+~ N^o_{\!A} ~=~ N_{\!A}\;,
\end{equation}
where $N_{\!A}$ is the dimension of the adjoint representation. For
instance, for $SU(N)$ one has $N^d_{\!A}$~$=$~$N-1$ and $N^o_{\!A}$~$=$~$%
N(N-1)$. Consequently, \cite{A4}, one can show that
\begin{eqnarray}
f^{iab} f^{iab} &=& N^d_{\!A} C_A ~~,~~ f^{abc} f^{abc} ~=~ \left[
N^o_{\!A} - 2 N^d_{\!A} \right] C_A ~~,~~ f^{icd} f^{jcd} ~=~ C_A
\delta^{ij}\;,
\nonumber \\
f^{acj} f^{bcj} &=& \frac{N^d_{\!A}}{N^o_{\!A}} C_A \delta^{ab}
~~,~~ f^{acd} f^{bcd} ~=~ \frac{[N^o_{\!A} - 2
N^d_{\!A}]}{N^o_{\!A}} C_A
\delta^{ab}\;,  \nonumber \\
f^{apq} f^{bpr} f^{cqr} &=& \frac{[N^o_{\!A} - 3
N^d_{\!A}]}{2N^o_{\!A}} C_A f^{abc} ~~,~~ f^{apq} f^{bpi} f^{cqi}
~=~ \frac{N^d_{\!A}}{2N^o_{\!A}} C_A
f^{abc}\;,  \nonumber \\
f^{ipq} f^{bpr} f^{cqr} &=& \frac{[N^o_{\!A} - 2
N^d_{\!A}]}{2N^o_{\!A}} C_A f^{ibc} ~~,~~ f^{ipq} f^{bpj} f^{cqj}
~=~ \frac{N^d_{\!A}}{N^o_{\!A}} C_A
f^{ibc}\;,  \nonumber \\
\mbox{Tr} \left( T^a T^b \right) &=& T_F \delta^{ab} ~~,~~ \mbox{Tr}
\left( T^a T^i \right) ~=~ 0 ~~,~~ \mbox{Tr} \left( T^i T^j \right)
~=~ T_F
\delta^{ij}\;,  \nonumber \\
T^i T^i &=& \frac{T_F}{N_{\!F}} N^d_{\!A} I ~~,~~ T^a T^a ~=~ \left[
C_F ~-~
\frac{T_F}{N_{\!F}} N^d_{\!A} \right] I\;,  \nonumber \\
T^b T^a T^b &=& \left[ C_F ~-~ \frac{C_A}{2} ~-~ \frac{T_F}{N_{\!F}}
N^d_{\!A} ~+~ \frac{C_A N^d_{\!A}}{2N^o_{\!A}} \right] T^a\;,  \nonumber \\
T^i T^a T^i &=& \left[ \frac{T_F}{N_{\!F}} N^d_{\!A} ~-~ \frac{C_A N^d_{\!A}%
}{2N^o_{\!A}} \right] T^a  \;,\nonumber \\
T^a T^i T^a &=& \left[ \frac{T_F}{N_{\!F}} N^o_{\!A} ~-~
\frac{C_A}{2} \right] T^i ~~,~~ T^j T^i T^j ~=~ \frac{T_F}{N_{\!F}}
N^d_{\!A} T^i\;,
\end{eqnarray}
where $N_{\!F}$ is the dimension of the fundamental representation
and satisfies
\begin{equation}
C_F N_{\!F} ~=~ \left[ N^o_{\!A} ~+~ N^d_{\!A} \right] T_F ~.
\end{equation}
For completeness we note that we have assumed that $f^{acd} f^{bcd}$
is proportional to $\delta^{ab}$ which is certainly true in $SU(2)$
and we have checked that it is valid in $SU(3)$, \cite{A4}.
Moreover, we have assumed that $T^i T^i$ is proportional to the unit
matrix, $I$, which is certainly true for $SU(N)$.

Before turning to the explicit discussion of how we renormalized the
interpolating gauge Lagrangian we define the anomalous dimensions of
the wave functions and parameters in terms of the renormalization
constants we will compute. We have
\begin{eqnarray}
\gamma_\phi(a) &=& \left[ \beta(a) \frac{\partial ~}{\partial a} ~+~
\sum_{n} \alpha_n \gamma_{\alpha_n}(a) \frac{\partial ~}{\partial
{\alpha_n}} \right] \ln Z_\phi \;,\nonumber
\end{eqnarray}
\begin{eqnarray}
\gamma_{\alpha_1}(a) &=& \left[ 1 - \alpha_1 \frac{\partial
~}{\partial \alpha_1} \ln Z_{\alpha_1} \right]^{-1} \left[ \, -~
\gamma_A(a) ~+~ \beta(a) \frac{\partial ~}{\partial a} ~+~
\sum_{n^\prime} \alpha_n \gamma_{\alpha_n}(a) \frac{\partial
~}{\partial {\alpha_n}} \right] \ln
Z_{\alpha_1}  \;,\nonumber \\
\gamma_{\alpha_2}(a) &=& -~ \left[ 1 + \alpha_2 \frac{\partial
~}{\partial
\alpha_2} \ln Z_{\alpha_2} \right]^{-1} \left[ \beta(a) \frac{\partial ~}{%
\partial a} ~+~ \sum_{n^\prime} \alpha_n \gamma_{\alpha_n}(a) \frac{\partial
~}{\partial {\alpha_n}} \right] \ln Z_{\alpha_2}  \;,\nonumber \\
\gamma_{\alpha_3}(a) &=& \left[ 1 - \alpha_3 \frac{\partial
~}{\partial \alpha_3} \ln Z_{\alpha_3} \right]^{-1} \left[ \, -~
\gamma_A(a) ~+~ \beta(a) \frac{\partial ~}{\partial a} ~+~
\sum_{n^\prime} \alpha_n \gamma_{\alpha_n}(a) \frac{\partial
~}{\partial {\alpha_n}} \right] \ln
Z_{\alpha_3}  \;,\nonumber \\
\gamma_{\alpha_4}(a) &=& -~ \left[ 1 + \alpha_4 \frac{\partial
~}{\partial
\alpha_4} \ln Z_{\alpha_4} \right]^{-1} \left[ \beta(a) \frac{\partial ~}{%
\partial a} ~+~ \sum_{n^\prime} \alpha_n \gamma_{\alpha_n}(a) \frac{\partial
~}{\partial {\alpha_n}} \right] \ln Z_{\alpha_4} \;, \nonumber \\
\gamma_{{\bar{\alpha}_5}}(a) &=& -~ \left[ 1 + {\bar{\alpha}_5} \frac{%
\partial ~}{\partial {\bar{\alpha}_5}} \ln Z_{{\bar{\alpha}_5}} \right]^{-1}
\left[ \beta(a) \frac{\partial ~}{\partial a} ~+~ \sum_{n^\prime}
\alpha_n
\gamma_{\alpha_n}(a) \frac{\partial ~}{\partial {\alpha_n}} \right] \ln Z_{{%
\bar{\alpha}_5}}\;,  \nonumber \\
\gamma_{\alpha_6}(a) &=& -~ \left[ 1 + \alpha_6 \frac{\partial
~}{\partial
\alpha_6} \ln Z_{\alpha_6} \right]^{-1} \left[ \beta(a) \frac{\partial ~}{%
\partial a} ~+~ \sum_{n^\prime} \alpha_n \gamma_{\alpha_n}(a) \frac{\partial
~}{\partial {\alpha_n}} \right] \ln Z_{\alpha_6}\;,
\end{eqnarray}
where $\sum_{n^\prime}$ means a sum over $1$ to $6$ but where the
term corresponding to the parameter associated with the left hand
side is omitted. We have defined $a$~$=$~$g^2/(16\pi^2)$. Moreover, $\phi$~$\in$~$\{A,A^i,\bar{c}c, \bar{c%
}^ic^i,\bar{c}^ic,\bar{c}c^i,\psi\}$ where we now use the notation
that a superscript $i$ on a field indicates that it is a diagonal
object. We have chosen to define the anomalous dimensions of the
ghost and anti-ghost renormalization constants in both sectors in a
non-standard way. This is because in the general interpolating gauge
to extract individual anomalous dimensions for each field itself
entails making an extra assumption. So, for
example, $Z_{\bar{c}c}$~$=$~$Z_{\bar{c}}^{\mbox{\small{$\frac{1}{2}$}}} Z_c^{%
\mbox{\small{$\frac{1}{2}$}}}$. We also note that our convention for
defining the relation between the bare and renormalized parameters for $%
\alpha_1$ and $\alpha_3$ is
\begin{equation}
\alpha_{1\,\mbox{\footnotesize{o}}} ~=~ Z^{-1}_{\alpha_1} Z_A \,
\alpha_1 ~~,~~ \alpha_{3\,\mbox{\footnotesize{o}}} ~=~
Z^{-1}_{\alpha_3} \tilde{Z}_A \, \alpha_3\;,
\end{equation}
where ${}_{\mbox{\footnotesize{o}}}$ denotes the bare quantity.

We now present the actual details of our renormalization. In order
to extract all the renormalization constants we have had to select a
particular set of Green's functions to renormalize. Since the order
in which we consider the particular interactions is crucial, it is
worth refering to the
actual Feynman rules given in the Appendix A. First, applying the \textsc{%
Mincer} algorithm, \cite{A5,A6}, to all the two-point functions we
have already discussed, it is straightforward to extract the wave
function renormalization constants and those for $\alpha _{1}$ and
$\alpha _{3}$ and encode them in the anomalous dimensions as
\begin{eqnarray}
\gamma _{A}(a) &=&\frac{1}{6N_{\!A}^{o}}\left[ N_{\!A}^{o}\left(
C_{A}(3\alpha _{1}-13)+8T_{F}N_{\!f}\right) \right.  \nonumber \\
&&\left. ~~~~~~~+~N_{\!A}^{d}\left( C_{A}(-3\alpha _{1}-9\alpha
_{2}+3\alpha
_{3}-3\alpha _{2}\alpha _{3})\right) \right] a~+~O(a^{2})\;,  \nonumber \\
\gamma _{A^{i}}(a) &=&\frac{1}{6}\left[ C_{A}(3\alpha _{1}\alpha
_{2}+3\alpha _{1}+9\alpha _{2}-13)+8T_{F}N_{\!f}\right]
a~+~O(a^{2})\;, \nonumber
\end{eqnarray}
\begin{eqnarray}
 \gamma _{\alpha _{1}}(a) &=&\frac{1}{12\alpha
_{1}N_{\!A}^{o}}\left[ N_{\!A}^{o}\left( C_{A}(-6\alpha
_{1}^2-6\alpha _{1}\alpha _{4}+26\alpha
_{1}-3\alpha _{4}^{2})-16\alpha _{1}T_{F}N_{\!f}\right) \right.  \nonumber \\
&&\left. ~~~~~~~~~~~~+~N_{\!A}^{d}C_{A}\left( 18\alpha
_{1}^{2}\alpha _{2}-12\alpha _{1}^{2}\bar{\alpha}_{5}+18\alpha
_{1}^{2}-6\alpha _{1}\alpha
_{2}^{2}\alpha _{3}-18\alpha _{1}\alpha _{2}^{2}\right. \right.  \nonumber \\
&&\left. \left. ~~~~~~~~~~~~~~~~~~~~~~~~~~~-~12\alpha _{1}\alpha
_{2}\alpha _{3}+6\alpha _{1}\alpha _{2}\alpha _{6}+18\alpha
_{1}\alpha _{2}-6\alpha
_{1}\alpha _{3}\right. \right.  \nonumber \\
&&\left. \left. ~~~~~~~~~~~~~~~~~~~~~~~~~~~+~12\alpha _{1}\alpha
_{4}-12\alpha _{1}\bar{\alpha}_{5}\alpha _{6}+6\alpha _{1}\alpha
_{6}-36\alpha _{2}^{2}+6\alpha _{4}^{2}\right) \right] a  \nonumber \\
&&+~O(a^{2})\;,  \nonumber \\
\gamma _{\alpha _{3}}(a) &=&\frac{1}{12\alpha _{3}}\left[
C_{A}\left( -6\alpha _{1}\alpha _{2}\alpha _{3}-6\alpha _{1}\alpha
_{3}-6\alpha _{2}\alpha _{3}\alpha _{6}-18\alpha _{2}\alpha
_{3}-6\alpha _{3}\alpha
_{6}+26\alpha _{3}-3\alpha _{6}^{2}\right) \right.  \nonumber \\
&&\left. ~~~~~~~~-~16\alpha _{3}T_{F}N_{\!f}\right] a~+~O(a^{2})\;,
\nonumber
\\
\gamma _{\bar{c}c}(a) &=&\frac{1}{4N_{\!A}^{o}}\left[
N_{\!A}^{o}C_{A}(\alpha _{1}-3)\right.  \nonumber \\
&&\left. ~~~~~~+~N_{\!A}^{d}C_{A}(3\alpha _{1}\alpha _{2}-2\alpha _{1}\bar{%
\alpha}_{5}+\alpha _{1}+\alpha _{2}^{2}\alpha _{3}-3\alpha
_{2}^{2}-2\alpha
_{2}\alpha _{3}\right.  \nonumber \\
&&\left. ~~~~~~~~~~~~~~~~~~~~+~\alpha _{2}\alpha _{6}+3\alpha
_{2}+\alpha
_{3}+\alpha _{6})\right] a~+~O(a^{2}) \;, \nonumber \\
\gamma _{\bar{c}^{i}c^{i}}(a) &=&\frac{C_{A}}{4}\left[ -\alpha
_{1}\alpha _{2}+2\alpha _{1}\bar{\alpha}_{5}-\alpha _{1}-\alpha
_{2}\alpha _{6}-3\alpha
_{2}-\alpha _{6}-3\right] a~+~O(a^{2}) \;, \nonumber \\
\gamma _{\psi }(a) &=&\frac{1}{N_{\!F}}\left[
N_{\!A}^{o}T_{F}(\alpha _{1}-\alpha _{3})+N_{\!F}C_{F}\alpha
_{3}\right] a~+~O(a^{2})\;,
\end{eqnarray}
where $N_{\!f}$ is the number of quarks. Due to our particular
definition of the renormalization of the gauge parameters $\alpha
_{1}$ and $\alpha _{3}$ we note that for completeness their
anomalous dimensions are
\begin{eqnarray}
\gamma _{A}(a)~+~\gamma _{\alpha _{1}}(a) &=&\frac{C_{A}}{4\alpha
_{1}N_{\!A}^{o}}\left[ N_{\!A}^{d}\left( 6\alpha _{1}^{2}\alpha
_{2}-4\alpha _{1}^{2}\bar{\alpha}_{5}+4\alpha _{1}^{2}-2\alpha
_{1}\alpha _{2}^{2}\alpha _{3}-6\alpha _{1}\alpha _{2}^{2}-6\alpha
_{1}\alpha _{2}\alpha _{3}\right.
\right.  \nonumber \\
&&\left. \left. ~~~~~~~~~~~~~~~~+2\alpha _{1}\alpha _{2}\alpha
_{6}+4\alpha _{1}\alpha _{4}-4\alpha _{1}\bar{\alpha}_{5}\alpha
_{6}+2\alpha _{1}\alpha
_{6}-12\alpha _{2}^{2}+2\alpha _{4}^{2}\right) \right.  \nonumber \\
&&\left. ~~~~~~~~~~~+~N_{\!A}^{o}\left( -2\alpha _{1}\alpha
_{4}-\alpha
_{4}^{2}\right) \right] a~+~O(a^{2}) \;, \nonumber \\
\gamma _{A^{i}}(a)~+~\gamma _{\alpha _{3}}(a) &=&-~\frac{\alpha _{6}C_{A}}{%
4\alpha _{3}}\left[ 2\alpha _{2}\alpha _{3}+2\alpha _{3}+\alpha
_{6}\right] a~+~O(a^{2})~.
\end{eqnarray}
The next stage is to consider the vertex renormalization. Equipped
with the diagonal and off-diagonal gluon and quark anomalous
dimensions we have renormalized the triple off-diagonal gluon, quark
off-diagonal gluon and quark diagional gluon three-point functions
and verified that the correct
one loop QCD $\beta $-function of \cite{A13} emerges, independent of \emph{%
all} six gauge fixing parameters, as
\begin{equation}
\beta (a)~=~-~\left[
\frac{11}{3}C_{A}-\frac{4}{3}T_{F}N_{\!f}\right] a^{2}~+~O(a^{3})~.
\end{equation}
This is a useful check on our procedure since it is known that the $\beta $%
-function is gauge independent in mass independent renormalization
schemes. For the four remaining renormalization constants those for
$\alpha _{2}$ and $\alpha _{4}$ are the easier to extract. From an
examination of the Feynman
rules then considering the respective vertices $\langle A^{i}A^{i}\bar{c}%
c\rangle $ and $\langle A\bar{c}c\rangle $ each parameter appears as
a factor and therefore their renormalization can be extracted by the
method of \cite{A12} given that $Z_{\bar{c}c}$ has been determined
from the off-diagonal ghost two-point function and $Z_{g}$ has first
been established as the usual value. In this instance for the
three-point function the momentum flow has been carefully chosen so
that there is no external momentum flowing into the $\bar{c}$ leg.
We find
\begin{eqnarray}
\gamma _{\alpha _{2}}(a) &=&\frac{C_{A}}{8\alpha
_{2}N_{\!A}^{o}}\left[ N_{\!A}^{o}\left( 2\alpha _{1}\alpha
_{2}^{2}+2\alpha _{1}\alpha
_{2}+6\alpha _{2}^{2}+6\alpha _{2}\right) \right.  \nonumber \\
&&\left. ~~~~~~~~~~+~N_{\!A}^{d}\left( 6\alpha _{1}\alpha
_{2}^{2}-2\alpha
_{1}\alpha _{2}\bar{\alpha}_{5}+8\alpha _{1}\alpha _{2}-2\alpha _{1}\bar{%
\alpha}_{5}+2\alpha _{1}\right. \right.  \nonumber \\
&&\left. \left. ~~~~~~~~~~~~~~~~~~~~-~6\alpha _{2}^{3}+\alpha
_{2}^{2}\alpha _{6}+2\alpha _{2}\alpha _{6}+6\alpha _{2}+\alpha
_{6}\right) \right]
a~+~O(a^{2}) \;, \nonumber \\
\gamma _{\alpha _{4}}(a) &=&\frac{1}{12\alpha _{4}N_{\!A}^{o}}\left[
N_{\!A}^{o}\left( C_{A}(3\alpha _{4}^{2}+26\alpha _{4})-16\alpha
_{4}T_{F}N_{\!f}\right) \right.  \nonumber \\
&&\left. ~~~~~~~~~~~~+~N_{\!A}^{d}C_{A}\left( 18\alpha _{1}\alpha
_{2}\alpha _{4}-12\alpha _{1}\alpha _{2}\alpha _{6}+12\alpha
_{1}\bar{\alpha}_{5}\alpha _{6}-12\alpha _{1}\alpha _{6}-18\alpha
_{2}^{2}\alpha _{4}\right. \right.
\nonumber \\
&&\left. \left. ~~~~~~~~~~~~~~~~~~~~~~~~~~~+~36\alpha
_{2}^{2}+6\alpha _{2}\alpha _{4}\alpha _{6}+18\alpha _{2}\alpha
_{4}-12\alpha
_{4}^{2}+6\alpha _{4}\alpha _{6}\right) \right] a~+~O(a^{2})~.  \nonumber \\
&&
\end{eqnarray}
To extract $Z_{\alpha _{6}}$ one first requires the renormalization
constant $Z_{\bar{c}^{i}c}$ since the vertices where $\alpha _{6}$
appears always involves the combination of fields $\bar{c}^{i}$ and
$c$. To do this we consider the vertex $\langle A\bar{c}^{i}c\rangle
$ and choose the momentum flow through the off-diagonal gluon leg to
be zero. From the Feynman rule this means that the tree term of the
Green's function does not depend on any parameters $\alpha _{i}$ and
therefore the only renormalization constant which is undetermined is
$Z_{\bar{c}^{i}c}$. Hence,
\begin{eqnarray}
\gamma _{\bar{c}^{i}c}(a) &=&\frac{1}{4N_{\!A}^{o}}\left[
N_{\!A}^{o}C_{A}(-\alpha _{1}\alpha _{2}+\alpha _{1}-\alpha
_{2}\alpha
_{6}-3\alpha _{2}+\alpha _{4}-\alpha _{6}-3)\right.  \nonumber \\
&&\left. ~~~~~~~+~N_{\!A}^{d}C_{A}(-2\alpha _{1}+2\alpha
_{3}-2\alpha _{4}+2\alpha _{6})\right] a~+~O(a^{2})~.
\end{eqnarray}
Equipped with this we can then determine $Z_{\alpha _{6}}$ in two
ways. The first is to renormalize the same vertex again but nullify
the momentum flow through the $\bar{c}^{i}$ leg which leaves a tree
term involving the as yet
undetermined $Z_{\alpha _{6}}$. An alternative way is to consider the $%
\langle AA^{i}\bar{c}^{i}c\rangle $ vertex given the diagonal and
off-diagonal gluon wave function renormalizations and $Z_{\alpha
_{2}}$ have already been determined. We have considered both and
deduced the \emph{same} renormalization constant for $Z_{\alpha
_{6}}$ for both Green's functions which is a useful check. We found
\begin{eqnarray}
\gamma _{\alpha _{6}}(a) &=&\frac{1}{12N_{\!A}^{o}}\left[
N_{\!A}^{o}\left( C_{A}(-3\alpha _{1}\alpha _{2}-3\alpha _{2}\alpha
_{6}-9\alpha _{2}+6\alpha
_{4}-3\alpha _{6}+26)-16T_{F}N_{\!f}\right) \right.  \nonumber \\
&&\left. ~~~~~~~~~+~N_{\!A}^{d}C_{A}\left( 9\alpha _{1}\alpha
_{2}+6\alpha _{1}\bar{\alpha}_{5}-9\alpha _{1}+3\alpha
_{2}^{2}\alpha _{3}-9\alpha _{2}^{2}+6\alpha _{2}\alpha _{3}+9\alpha
_{2}\alpha _{6}\right. \right.
\nonumber \\
&&\left. \left. ~~~~~~~~~~~~~~~~~~~~~~~~+~9\alpha _{2}+3\alpha
_{3}-12\alpha _{4}+9\alpha _{6}\right) \right] a~+~O(a^{2})~.
\end{eqnarray}
Finally, we follow a similar procedure to extract
$Z_{\bar{\alpha}_{5}}$. However, given we have just found $Z_{\alpha
_{6}}$ by considering the vertex $\langle
\bar{c}c\bar{c}^{i}c^{i}\rangle $ then $Z_{\bar{\alpha}_{5}}$ can be
extracting immediately given that the wave function renormalization
for this vertex can be written into the already determined combination $Z_{%
\bar{c}c}$ and $Z_{\bar{c}^{i}c^{i}}$. This gives
\begin{eqnarray}
\gamma _{\bar{\alpha}_{5}}(a) &=&\frac{C_{A}}{4\alpha
_{1}N_{\!A}^{o}}\left[ N_{\!A}^{o}\left( 2\alpha
_{1}^{2}\bar{\alpha}_{5}-2\alpha _{1}^{2}+\alpha
_{4}^{2}\right) \right.  \nonumber \\
&&\left. ~~~~~~~~~~+~N_{\!A}^{d}\left( -4\alpha _{1}\alpha
_{2}\alpha _{6}+~4\alpha _{1}\bar{\alpha}_{5}\alpha _{6}-4\alpha
_{1}\alpha _{6}+12\alpha _{2}^{2}-2\alpha _{4}^{2}\right) \right]
a~+~O(a^{2})\,.~~~~~~~~
\end{eqnarray}
However, to obtain the complete set of renormalization constants we
still need to deduce $Z_{\bar{c}c^{i}}$. Again this can be extracted
from two
vertices which are $\langle A\bar{c}c^{i}\rangle $ and $\langle AA^{i}\bar{c}%
c^{i}\rangle $ where there is no momentum flowing through the
$c^{i}$ leg in the former case. We have renormalized both Green's
functions to ensure that the same renormalization constant emerges
and found in \emph{both} cases
\begin{eqnarray}
\gamma _{\bar{c}c^{i}}(a) &=&\frac{1}{4N_{\!A}^{o}}\left[
N_{\!A}^{o}C_{A}(2\alpha _{1}\bar{\alpha}_{5}-\alpha _{1}-\alpha
_{4}-3)\right.  \nonumber \\
&&\left. ~~~~~~~+~N_{\!A}^{d}C_{A}(3\alpha _{1}\alpha _{2}-2\alpha _{1}\bar{%
\alpha}_{5}+3\alpha _{1}+\alpha _{2}^{2}\alpha _{3}-3\alpha
_{2}^{2}-2\alpha
_{2}\alpha _{3}\right.  \nonumber \\
&&\left. ~~~~~~~~~~~~~~~~~~~~~~+~\alpha _{2}\alpha _{6}+3\alpha
_{2}-\alpha _{3}+2\alpha _{4}-\alpha _{6})\right] a~+~O(a^{2})\;,
\end{eqnarray}
which is a useful internal check.

Having produced the anomalous dimensions for the general
interpolating gauge we are now in a position to take the various
limits of the four special cases discussed above in order to check
with the corresponding known anomalous dimensions for these gauges.
In comparing with known anomalous dimensions any agreement we
indicate is for all the anomalous dimensions
except for the interpolating parameters $\alpha_2$, $\alpha_4$, $\bar{\alpha}%
_5$ and $\alpha_6$. For the linear covariant gauge case we have
\begin{eqnarray}
\gamma_A(a) &=&\gamma_{A^i}(a)= \frac{1}{6} \left( C_A ( 3 \alpha -
13 ) + 8 T_F N_{\!f}
\right) a ~+~ O(a^2) \;, \nonumber \\
\gamma_{\alpha_1}(a) &=&\gamma_{\alpha_3}(a)= -~ \frac{1}{6} \left(
C_A ( 3 \alpha - 13 ) + 8 T_F
N_{\!f} \right) a ~+~ O(a^2)\;,  \nonumber \\
\gamma_{\alpha_2}(a) &=& \frac{C_A}{4N^o_{\!A}} \left[ N^o_{\!A}
\left( \alpha + 3 \right) + 3 N^d_{\!A} \left( \alpha + 1 \right)
\right] a ~+~
O(a^2) \;, \nonumber \\
\gamma_{\alpha_4}(a) &=&\gamma_{\alpha_6}(a)= \frac{1}{6} \left( 13
C_A - 8 T_F N_{\!f} \right) a
~+~ O(a^2)\;,  \nonumber \\
\gamma_{\bar{\alpha}_5}(a) &=& O(a^2)  \;,\nonumber \\
\gamma_{\bar{c}c}(a) &=&\gamma_{\bar{c}^ic^i}(a)=\gamma_{\bar{c}c^i}(a)= \gamma_{\bar{c}^ic}(a)=\frac{(\alpha-3)C_A}{4} a ~+~ O(a^2)\;,  \nonumber \\
\gamma_\psi(a) &=& \alpha C_F a ~+~ O(a^2)\;,
\end{eqnarray}
where for $\gamma_{\alpha_4}(a)$ we have taken the limit $\alpha_2$~$%
\rightarrow$~$0$ before setting $\alpha_4$~$=$~$0$. These
expressions are in agreement with the well established linear
covariant gauge anomalous dimensions. For the Curci-Ferrari gauge we
have
\begin{eqnarray}
\gamma_A(a) &=&\gamma_{A^i}(a)= \frac{1}{6} \left( C_A ( 3 \alpha -
13 ) + 8 T_F N_{\!f} \right) a ~+~ O(a^2)\;,  \nonumber
\\
 \gamma_{\alpha_1}(a) &=&\gamma_{\alpha_3}(a)=\gamma_{\alpha_4}(a)=\gamma_{\alpha_6}(a)= -~
\frac{1}{12} \left( C_A ( 3 \alpha - 26 ) + 16
T_F N_{\!f} \right) a ~+~ O(a^2)\;,  \nonumber \\
\gamma_{\alpha_2}(a) &=& \frac{C_A}{8N^o_{\!A}} \left[ N^o_{\!A}
\left( 2 \alpha + 6 \right) + N^d_{\!A} \left( 5 \alpha + 6 \right)
\right] a ~+~
O(a^2) \;, \nonumber \\
\gamma_{\bar{\alpha}_5}(a) &=& O(a^2)\;,  \nonumber \\
\gamma_{\bar{c}c}(a) &=&\gamma_{\bar{c}^ic^i}(a)= \gamma_{\bar{c}c^i}(a)=\gamma_{\bar{c}^ic}(a)=\frac{(\alpha-3)C_A}{4} a ~+~ O(a^2)\;,  \nonumber \\
\gamma_\psi(a) &=& \alpha C_F a ~+~ O(a^2)\;,
\end{eqnarray}
which are also in agreement with \cite{A14,A15,A16,A17}. The
equality of the four ghost wave function anomalous dimensions in
this and the previous case is consistent with each of the individual
wave function ghost anomalous dimensions being equal and equivalent
to the value given. For the first of the MAG cases taking the limits
with the values in Table $3$ we find
\begin{eqnarray}
\gamma_A(a) &=& \frac{1}{6N^o_{\!A}} \left[ N^o_{\!A} \left( C_A ( 3
\alpha - 13 ) + 8 T_F N_{\!f} \right) + N^d_{\!A} C_A ( - 3 \alpha +
9 ) \right] a
~+~ O(a^2)\;,  \nonumber \\
\gamma_{A^i}(a) &=& -~ \frac{1}{3} \left( 11 C_A - 4 T_F N_{\!f}
\right) a
~+~ O(a^2)\;,  \nonumber \\
\gamma_{\alpha_1}(a) &=&\gamma_{\alpha_4}(a)= \frac{1}{12\alpha
N^o_{\!A}} \left[ N^o_{\!A} \left( C_A ( - 3 \alpha^2 + 26 \alpha )
- 16 \alpha T_F N_{\!f} \right)
\right.  \nonumber \\
&& \left. ~~~~~~~~~~~ +~ N^d_{\!A} C_A ( - 6 \alpha^2 - 36 \alpha -
36 )
\right] a ~+~ O(a^2)\;,  \nonumber \\
\gamma_{\alpha_2}(a) &=& O(a^2)\;,  \nonumber \\
\gamma_{\alpha_3}(a) &=& \frac{1}{3} \left( 11 C_A - 4 T_F N_{\!f}
\right) a
~+~ O(a^2) \;, \nonumber \\
\gamma_{\bar{\alpha}_5}(a) &=& \frac{C_A}{4\alpha N^o_{\!A}} \left(
N^d_{\!A} \left( 12 - 2 \alpha^2 \right) - \alpha^2 N^o_{\!A}
\right) a ~+~
O(a^2) \;, \nonumber \\
\gamma_{\alpha_6}(a) &=& \frac{1}{12N^o_{\!A}} \left[ N^o_{\!A}
\left( C_A ( - 3 \alpha + 35 ) - 16 T_F N_{\!f} \right) + N^d_{\!A}
C_A ( - 6 \alpha - 18
) \right] a ~+~ O(a^2)\;,  \nonumber \\
\gamma_{\bar{c}c}(a) &=& \frac{C_A}{4N^o_{\!A}} \left( N^d_{\!A}
\left( - 6 - 2 \alpha \right) + N^o_{\!A} \left( \alpha - 3 \right)
\right) a ~+~ O(a^2)\;,\nonumber\\
\gamma_{\bar{c}^ic^i}(a) &=& O(a^2) \;, \nonumber \\
\gamma_{\bar{c}c^i}(a) &=& \frac{C_A}{4N^o_{\!A}} \left( N^d_{\!A}
\left( -
6 - 2 \alpha \right) - 3 N^o_{\!A} \right) a ~+~ O(a^2) \;, \nonumber \\
\gamma_{\bar{c}^ic}(a) &=& \frac{\alpha C_A}{4} a ~+~ O(a^2) \;, \nonumber \\
\gamma_\psi(a) &=& \frac{\alpha N^o_{\!A} T_F}{N_{\!F}} a ~+~
O(a^2)\;.
\end{eqnarray}
These are in agreement with the one loop arbitrary color group
anomalous dimensions recorded in \cite{A4}. Although unlike the
previous two cases the combination of the ghost wave function
anomalous dimensions are not equal, their values are consistent with
the values given in \cite{A4}. Finally, we quote the values for the
parameters stated in Table $4$ are
\begin{eqnarray}
\gamma_A(a) &=& \frac{1}{6N^o_{\!A}} \left[ N^o_{\!A} \left( C_A ( 3
\alpha - 13 ) + 8 T_F N_{\!f} \right) - 3 \alpha C_A N^d_{\!A}
\right] a ~+~ O(a^2)\;,
\nonumber \\
\gamma_{A^i}(a) &=& \frac{1}{6} \left( ( 3 \alpha - 13 ) C_A + 8 T_F
N_{\!f}
\right) a ~+~ O(a^2)\;,  \nonumber \\
\gamma_{\alpha_1}(a) &=&\gamma_{\alpha_4}(a)= \frac{1}{12N^o_{\!A}}
\left[ N^o_{\!A} \left( C_A ( - 3 \alpha + 26 ) - 16 T_F N_{\!f}
\right) + 12 \alpha N^d_{\!A} C_A \right] a ~+~ O(a^2) \;,
\nonumber\\
 \gamma_{\alpha_3}(a) &=& \frac{1}{6} \left( ( 13 - 3 \alpha ) C_A
- 8 T_F N_{\!f} \right) a ~+~ O(a^2)\;,  \nonumber\\
 \gamma_{\bar{\alpha}_5}(a) &=& -~ \frac{\alpha C_A}{4N^o_{\!A}}
\left( 2
N^d_{\!A} + N^o_{\!A} \right) a ~+~ O(a^2) \;, \nonumber \\
\gamma_{\alpha_6}(a) &=& \frac{1}{12N^o_{\!A}} \left[ N^o_{\!A}
\left( C_A ( - 6 \alpha + 26 ) - 16 T_F N_{\!f} \right) + 3 \alpha
N^d_{\!A} C_A \right]
a ~+~ O(a^2) \;, \nonumber \\
\gamma_{\bar{c}c}(a) &=& \frac{C_A}{4N^o_{\!A}} \left( \alpha
N^d_{\!A} +
N^o_{\!A} \left( \alpha - 3 \right) \right) a ~+~ O(a^2) \;, \nonumber \\
\gamma_{\bar{c}^ic^i}(a) &=& -~ \frac{(\alpha+3)C_A}{4} a ~+~ O(a^2)
\;,\nonumber \\
\gamma_{\bar{c}c^i}(a) &=& \frac{C_A}{4N^o_{\!A}} \left( \alpha
N^d_{\!A} -
3 N^o_{\!A} \right) a ~+~ O(a^2)\;,  \nonumber \\
\gamma_{\bar{c}^ic}(a) &=& -~ \frac{3 C_A}{4} a ~+~ O(a^2)\;,  \nonumber \\
\gamma_\psi(a) &=& \frac{\alpha N^o_{\!A} T_F}{N_{\!F}} a ~+~
O(a^2)\;,\label{mmagnew}
\end{eqnarray}
where $\gamma_{\alpha_2}(a)$ is singular as $\alpha_2$ $\rightarrow$
$0$. However, this singularity does not affect the anomalous
dimensions of the gauge parameters entering the modified maximal
Abelian gauge. We regard the results (\ref{mmagnew}) as new since as
far as we are aware there are no similar computations in this
specific gauge.

\sect{Conclusion.} We have constructed a six parameter family of
covariant gauge fixings, which allows us to interpolate between
various familiar gauges: the Landau, the linear covariant, the
Curci-Ferrari and the (modified) maximal Abelian gauges. Using the
algebraic renormalization formalism, we have been able to prove the
renormalizability of the proposed gauge to any order of perturbation
theory. In addition, we have also computed explicitly every
anomalous dimension at one-loop order in the $\MSbar$ scheme,
confirming that the obtained results reduce to the already known
values in the various special cases, if already known. The gauge
parameter independence of the gauge beta function was confirmed.

We end by noticing that the concept of an interpolating gauge may
also find potential application in lattice numerical simulations.
More precisely, requiring that the functional
\begin{equation}\label{interpolk}
  \mathcal{R}_k[A]=(VT)^{-1}\int d^4 x \left(A^a_{\mu}A^{\mu a} + k A^i_\mu
A^{i \mu} \right)\;,
\end{equation}
with $k$ the interpolating parameter, attains a minimum with respect
to local gauge transformations, leads to the MAG for $k=0$ and to
the Landau gauge for $k=1$. This could be useful in order to study
the presence, in the MAG, of power corrections in $\frac{1}{q^2}$ to
the strong coupling constant. Such power corrections are by now
quite well established in the Landau gauge,
\cite{Boucaud:2001st,RuizArriola:2004en,Furui:2005bu,Boucaud:2005rm}.
As it turns out that one can interpolate from the Landau gauge to
the MAG, one could therefore expect similar corrections in the MAG.
One might also hope to obtain some information on the relevance of
the Gribov copies in the MAG.

\section*{Acknowledgments.}
The Conselho Nacional de Desenvolvimento Cient\'{i}fico e
Tecnol\'{o}gico (CNPq-Brazil), the Faperj, Funda{\c{c}}{\~{a}}o de
Amparo {\`{a}} Pesquisa do Estado do Rio de Janeiro, the SR2-UERJ
and the Coordena{\c{c}}{\~{a}}o de Aperfei{\c{c}}oamento de Pessoal
de N{\'{i}}vel Superior (CAPES) are gratefully acknowledged for
financial support.

\appendix
\sect{Feynman rules.} In this appendix we list the explicit Feynman
rules we used in momentum space. For the propagators we have
\begin{eqnarray}
\langle A_{\mu }^{a}(p)A_{\nu }^{b}(-p)\rangle &=&-~\frac{\delta ^{ab}}{p^{2}%
}\left[ \eta _{\mu \nu }~-~(1-\alpha _{1})\frac{p_{\mu }p_{\nu }}{p^{2}}%
\right]\;,  \nonumber \\
\langle A_{\mu }^{i}(p)A_{\nu }^{j}(-p)\rangle &=&-~\frac{\delta ^{ij}}{p^{2}%
}\left[ \eta _{\mu \nu }~-~(1-\alpha _{3})\frac{p_{\mu }p_{\nu }}{p^{2}}%
\right]\;,  \nonumber \\
\langle c^{a}(p){\bar{c}}^{b}(-p)\rangle &=&\frac{\delta ^{ab}}{p^{2}}%
~~,~~\langle c^{i}(p){\bar{c}}^{j}(-p)\rangle ~=~\frac{\delta ^{ij}}{p^{2}}%
~~,~~\langle \psi (p){\bar{\psi}}(-p)\rangle
~=~\frac{p\!\!\!/}{p^{2}}\;,
\end{eqnarray}
where $p$ is the momentum. The non-zero $3$- and $4$-point vertices are%
\footnote{%
We note that in \cite{A4} the corresponding Feynman rules for the
four-point corrections did not all have the correct coupling
constant factor of $g^2$ and the $\langle A A A^i \rangle$ vertex
was inadvertently omitted.}
\begin{eqnarray}
\langle A_{\mu }^{a}(p_{1})\bar{\psi}(p_{2})\psi (p_{3})\rangle
&=&gT^{a}\gamma _{\mu } \;, \nonumber \\
\langle A_{\mu }^{i}(p_{1})\bar{\psi}(p_{2})\psi (p_{3})\rangle
&=&gT^{i}\gamma _{\mu } \;, \nonumber \\
\langle A_{\mu }^{a}(p_{1})\bar{c}^{b}(p_{2})c^{c}(p_{3})\rangle
&=&-~igf^{abc}\left( -\left( 1+\frac{\alpha _{4}}{2\alpha
_{1}}\right)
p_{1}-p_{3}\right) _{\mu } \;, \nonumber \\
\langle A_{\mu }^{a}(p_{1})\bar{c}^{b}(p_{2})c^{k}(p_{3})\rangle
&=&-~igf^{abk}\left( -\bar{\alpha}_{5}p_{1}-(1+\alpha
_{2})p_{3}\right)
_{\mu } \;, \nonumber \\
\langle A_{\mu }^{a}(p_{1})\bar{c}^{j}(p_{2})c^{c}(p_{3})\rangle
&=&-~igf^{acj}\left( \left( 1+\frac{\alpha _{6}}{2\alpha
_{1}}\right)
p_{1}+p_{3}\right) _{\mu }\;,  \nonumber \\
\langle A_{\mu }^{i}(p_{1})\bar{c}^{b}(p_{2})c^{c}(p_{3})\rangle
&=&-~igf^{bci}\left( -\left( 1+\frac{\alpha _{6}}{2\alpha
_{3}}\right)
p_{1}-(1-\alpha _{2})p_{3}\right) _{\mu } \;, \nonumber \\
\langle A_{\mu }^{a}(p_{1})A_{\nu }^{b}(p_{2})A_{\sigma
}^{c}(p_{3})\rangle &=&igf^{abc}\left( \eta _{\nu \sigma
}(p_{2}-p_{3})_{\mu }+\eta _{\sigma \mu }(p_{3}-p_{1})_{\nu }+\eta
_{\mu \nu }(p_{1}-p_{2})_{\sigma }\right)\;,
\nonumber \\
\langle A_{\mu }^{a}(p_{1})A_{\nu }^{b}(p_{2})A_{\sigma
}^{k}(p_{3})\rangle &=&igf^{abk}\left( \eta _{\nu \sigma }\left(
p_{2}-\frac{\alpha _{2}}{\alpha _{1}}p_{1}-p_{3}\right) _{\mu }+\eta
_{\sigma \mu }\left( p_{3}+\frac{\alpha
_{2}}{\alpha _{1}}p_{2}-p_{1}\right) _{\nu }\right.  \nonumber \\
&&\left. ~~~~~~~~~~+\eta _{\mu \nu }(p_{1}-p_{2})_{\sigma }\right)
\nonumber\;,
\\
\langle A_{\mu }^{a}(p_{1})A_{\nu }^{b}(p_{2})A_{\sigma
}^{c}(p_{3})A_{\rho }^{d}(p_{4})\rangle &=&g^{2}\left[
f_{d}^{abcd}\left( \eta _{\mu \sigma }\eta _{\nu \rho }-\eta _{\mu
\rho }\eta _{\nu \sigma }\right) +f_{d}^{acbd}\left( \eta _{\mu \nu
}\eta _{\sigma \rho }-\eta _{\mu \rho
}\eta _{\nu \sigma }\right) \right.  \nonumber \\
&&\left. ~~~~+f_{d}^{adbc}\left( \eta _{\mu \nu }\eta _{\sigma \rho
}-\eta _{\mu \sigma }\eta _{\nu \rho }\right) +f_{o}^{abcd}\left(
\eta _{\mu \sigma }\eta _{\nu \rho }-\eta _{\mu \rho }\eta _{\nu
\sigma }\right) \right.
\nonumber \\
&&\left. ~~~~+f_{o}^{acbd}\left( \eta _{\mu \nu }\eta _{\sigma \rho
}-\eta _{\mu \rho }\eta _{\nu \sigma }\right) +f_{o}^{adbc}\left(
\eta _{\mu \nu }\eta _{\sigma \rho }-\eta _{\mu \sigma }\eta _{\nu
\rho }\right) \right] \;,\nonumber
\end{eqnarray}
\begin{eqnarray}
\langle A_{\mu }^{a}(p_{1})A_{\nu }^{b}(p_{2})A_{\sigma
}^{c}(p_{3})A_{\rho }^{l}(p_{4})\rangle &=&-~g^{2}\left(
f_{o}^{abcl}\left( -\eta _{\mu \sigma }\eta _{\nu \rho }+\eta _{\mu
\rho }\eta _{\nu \sigma }\right) +f_{o}^{acbl}\left( -\eta _{\mu \nu
}\eta _{\sigma \rho }+\eta _{\mu \rho
}\eta _{\nu \sigma }\right) \right.  \nonumber \\
&&\left. ~~~~~~~+f_{o}^{albc}\left( -\eta _{\mu \nu }\eta _{\sigma
\rho
}+\eta _{\mu \sigma }\eta _{\nu \rho }\right) \right)\;,  \nonumber \\
\langle A_{\mu }^{a}(p_{1})A_{\nu }^{b}(p_{2})A_{\sigma
}^{k}(p_{3})A_{\rho }^{l}(p_{4})\rangle &=&-~g^{2}\left(
f_{o}^{akbl}\left( -\eta _{\mu \nu }\eta _{\sigma \rho
}-\frac{\alpha _{2}^{2}}{\alpha _{1}}\eta _{\mu \sigma }\eta _{\nu
\rho }+\eta _{\mu \sigma }\eta _{\nu \rho }+\eta _{\mu \rho
}\eta _{\nu \sigma }\right) \right.  \nonumber \\
&&\left. ~~~~~~~~~+f_{o}^{albk}\left( -\eta _{\mu \nu }\eta _{\sigma
\rho
}+\eta _{\mu \sigma }\eta _{\nu \rho }-\frac{\alpha _{2}^{2}}{\alpha _{1}}%
\eta _{\mu \rho }\eta _{\nu \sigma }\right) \right)  \;,\nonumber \\
\langle A_{\mu }^{a}(p_{1})A_{\nu }^{b}(p_{2})\bar{c}^{c}(p_{3})c^{d}(p_{4})%
\rangle &=&-~g^{2}\left( f_{d}^{acbd}\alpha _{2}\eta _{\mu \nu
}+f_{d}^{bcad}\alpha _{2}\eta _{\mu \nu }\right)  \;,\nonumber \\
\langle A_{\mu }^{a}(p_{1})A_{\nu }^{j}(p_{2})\bar{c}^{c}(p_{3})c^{d}(p_{4})%
\rangle &=&-~g^{2}\left( f_{o}^{adcj}\alpha _{2}\eta _{\mu \nu }+f_{o}^{ajcd}%
\frac{\alpha _{2}\alpha _{4}}{2\alpha _{1}}\eta _{\mu \nu
}\right)\;, \nonumber
\\
\langle A_{\mu }^{a}(p_{1})A_{\nu }^{j}(p_{2})\bar{c}^{c}(p_{3})c^{l}(p_{4})%
\rangle &=&-~g^{2}\left( f_{o}^{ajcl}\frac{\alpha _{2}\alpha
_{5}}{2\alpha _{1}}\eta _{\mu \nu }+f_{o}^{alcj}\alpha _{2}\eta
_{\mu \nu }\right)\;,
\nonumber \\
\langle A_{\mu }^{a}(p_{1})A_{\nu }^{j}(p_{2})\bar{c}^{k}(p_{3})c^{d}(p_{4})%
\rangle &=&-~g^{2}\left( -f_{o}^{ajdk}\frac{\alpha _{2}\alpha
_{6}}{2\alpha
_{1}}\eta _{\mu \nu }\right)\;,  \nonumber \\
\langle A_{\mu }^{i}(p_{1})A_{\nu }^{j}(p_{2})\bar{c}^{c}(p_{3})c^{d}(p_{4})%
\rangle &=&-~g^{2}\left( -f_{o}^{cidj}\alpha _{2}\eta _{\mu \nu
}-f_{o}^{cjdi}\alpha _{2}\eta _{\mu \nu }\right)  \;,\nonumber \\
\langle
\bar{c}^{a}(p_{1})c^{b}(p_{2})\bar{c}^{c}(p_{3})c^{d}(p_{4})\rangle
&=&-~g^{2}\left( -\frac{\alpha _{6}^{2}}{4\alpha _{3}}f_{d}^{abcd}-\frac{%
\alpha _{5}}{2}f_{d}^{acbd}-\frac{\alpha _{6}^{2}}{4\alpha _{3}}%
f_{d}^{adbc}\right.  \nonumber \\
&&\left. ~~~~~~~~~-\frac{\alpha _{4}^{2}}{4\alpha _{1}}f_{o}^{abcd}-\frac{%
\alpha _{4}}{2}f_{o}^{acbd}-\frac{\alpha _{4}^{2}}{4\alpha _{1}}%
f_{o}^{adbc}\right) \;, \nonumber\\
 \langle
\bar{c}^{a}(p_{1})c^{b}(p_{2})\bar{c}^{c}(p_{3})c^{l}(p_{4})\rangle
&=&-~g^{2}\left( -\frac{\alpha _{4}\alpha _{5}}{4\alpha _{1}}f_{o}^{abcl}-%
\frac{\alpha _{4}}{2}f_{o}^{acbl}-\frac{\alpha _{4}\alpha _{5}}{4\alpha _{1}}%
f_{o}^{albc}\right) \;, \nonumber \\\langle
\bar{c}^{a}(p_{1})c^{b}(p_{2})\bar{c}^{k}(p_{3})c^{d}(p_{4})\rangle
&=&-~g^{2}\left( \frac{\alpha _{4}\alpha _{6}}{4\alpha _{1}}f_{o}^{abdk}-%
\frac{\alpha _{4}\alpha _{6}}{4\alpha _{1}}f_{o}^{adbk}-\frac{\alpha _{6}}{2}%
f_{o}^{akbd}\right)\;,  \nonumber \\
\langle
\bar{c}^{a}(p_{1})c^{b}(p_{2})\bar{c}^{k}(p_{3})c^{l}(p_{4})\rangle
&=&-~g^{2}\left( -\frac{\alpha _{6}}{2}f_{o}^{akbl}-\frac{\alpha
_{5}\alpha _{6}}{4\alpha _{1}}f_{o}^{albk}\right)\;,\nonumber\\
\langle
\bar{c}^{a}(p_{1})c^{j}(p_{2})\bar{c}^{c}(p_{3})c^{l}(p_{4})\rangle
&=&0 \;, \nonumber \\
\langle
\bar{c}^{i}(p_{1})c^{b}(p_{2})\bar{c}^{k}(p_{3})c^{d}(p_{4})\rangle
&=&0\;,  \label{feynrule}
\end{eqnarray}
where the momentum, $p_{i}$, flow into each vertex and we have
introduced the more compact notation, \cite{A4},
\begin{equation}
f_{d}^{ABCD}~=~f^{iAB}f^{iCD}~~,~~f_{o}^{ABCD}~=~f^{eAB}f^{eCD}\;,
\end{equation}
for the four-point interactions with $i$, $j$, $k$ and $l$ denoting
indices of objects in the group centre. The Feynman rules for the
remaining possible combinations of diagonal and off-diagonal fields
are trivially zero since they would generate group factors involving
$f^{ijc}$ or $f^{ijk}$ which are both zero from the Lie algebra.

\end{document}